\documentclass[epj]{svjour}
\usepackage{amsmath}
\usepackage{graphics}
\usepackage{graphicx}
\usepackage{tabularx}
\usepackage{psfrag}
\usepackage{comment}

\usepackage{url}
\usepackage{epsfig}
\usepackage{array}
\usepackage{hyperref}
\usepackage[english]{babel}

\usepackage{epsfig}
\usepackage{cite}
\usepackage{amssymb}
\usepackage{latexsym}
\usepackage{indentfirst}
\usepackage{color}

\begin{document}
%
%
\title{Nucleon and nuclear structure through dilepton production}

\author{
I.V.~Anikin\inst{1}, 
N.~Baltzell\inst{2}, 
M.~Boer\inst{3}, 
R.~Boussarie\inst{4}, 
V.M.~Braun\inst{5}, 
S.J.~Brodsky\inst{6}, 
A.~Camsonne\inst{2}, 
W.C.~Chang\inst{7}, 
L.~Colaneri\inst{8}, 
S.~Dobbs\inst{9,10}, 
A.V.~Efremov\inst{1}, 
K.~Gnanvo\inst{11}, 
O.~Gryniuk\inst{12}, 
M.~Guidal\inst{8}, 
V.~Guzey\inst{13}, 
C.E.~Hyde\inst{14}, 
Y.~Ilieva\inst{15}, 
S.~Joosten\inst{16},
P.~Kroll\inst{17}, 
K.~Kumeri\v{c}ki\inst{18}, 
Z.-E.~Meziani\inst{16}, 
D.~M\"uller\inst{18}, 
K.M.~Semenov-Tian-Shansky\inst{13}, 
S.~Stepanyan\inst{2}, 
L.~Szymanowski\inst{19},
V.~Tadevosyan\inst{20},
O.V.~Teryaev\inst{1},
M. Vanderhaeghen\inst{12}, 
E.~Voutier\inst{8,}
\thanks{\emph{Corresponding author:} voutier@ipno.in2p3.fr},
J.~Wagner\inst{19}, 
C.~Weiss\inst{2}, 
Z.W.~Zhao\inst{21}
}                     
\institute{
JINR,Dubna, Russia \and 
Thomas Jefferson National Accelerator Facility, Newport News, VA, USA \and 
Los Alamos National Laboratory, Los Alamos, NM, USA \and
Institute of Nuclear Physics, Polish Academy of Sciences, Krak\'o w, Poland \and 
Universit\"at Regensburg, Regensburg, Germany \and
SLAC National Accelerator Laboratory, Stanford University, Stanford, CA, USA \and 
Institute of Physics, Academia Sinica, Taipei 11529, Taiwan \and
Institut de Physique Nucl\'eaire, Universit\'es Paris-Sud \& Paris-Saclay, CNRS/IN2P3, Orsay, France \and
Northwestern University, Evanston, IL  60208, USA \and
Florida State University, Tallahassee, FL, USA \and
University of Virginia, Charlottesville, VA, USA \and
Institute of Nuclear Physics Johann-Joachim-Becher, 55128 Mainz, Germany \and
National Research Centre Kurchatov Institute, Petersburg Nuclear Physics Institute, Gatchina, Russia \and  
Old Dominion University, Norfolk, VA, USA \and 
University of South Carolina, Columbia, SC, USA \and 
Temple University, Philadelphia, PA, USA \and  
Universit\"at Wuppertal, Wuppertal, Germany \and 
University of Zagreb, Croatia \and
National Centre for Nuclear Research, Warsaw, Poland \and 
A.I.~Alikhanian National Science Laboratory, Yerevan, Armenia \and 
Duke University, Durham, NC, USA
}

\date{Received: date / Revised version: date} 

\abstract{ %
Transverse momentum distributions and generalized parton distributions provide a comprehensive framework for the 
three-dimensional imaging of the nucleon and the nucleus experimentally using deeply virtual semi-exclusive and 
exclusive processes. The advent of combined high luminosity facilities and large acceptance detector capabilities 
enables experimental investigation of the partonic structure of hadrons with time-like virtual probes, in complement 
to the rich on-going space-like virtual probe program. The merits and benefits of the dilepton production channel 
for nuclear structure studies are discussed within the context of the International Workshop on Nucleon and 
Nuclear Structure through Dilepton Production taking place at the European Center for Theoretical Studies in Nuclear 
Physics and Related Areas (ECT$^{\star}$) of Trento. Particularly, the double deeply virtual Compton scattering, 
the time-like Compton scattering, the deeply virtual meson production, and the Drell-Yan processes are reviewed 
and a strategy for high impact experimental measurements is proposed. 
} 

\authorrunning{I.V.~Anikin et al.}
\titlerunning{Dilepton production}

\maketitle

\hyphenation{az-mu-thal-ly}
\hyphenation{Phys-sics}

%
%
\section{Introduction}

The comprehensive understanding of the nucleon and nuclear structure and dynamics in terms of the elementary bricks of matter, the quarks and the gluons or more generally the partons, is the essential quest of modern hadronic physics. After decades of theoretical and experimental efforts in elastic and deep inelastic lepton scatterings, a new formulation of the hadron structure~\cite{{Mue94},{Ji97},{Rad97}} emerges that builds any hadron from the correlations between its inner partons. This powerful approach provides not only a tomographic image~\cite{{Bur00},{Ral02},{Die02}} of the nucleon and the nucleus, but also a comprehensive description of their internal dynamics which expresses for example the origin of the nucleon spin~\cite{Ji97} or the distribution of the strong forces between hadron constituents~\cite{Pol03}. These perspectives, developed within the context of the formalism of the Generalized Parton Distributions (GPDs) and the Transverse Momentum Distributions (TMDs), motivated an intense experimental activity at many centers (DESY, JLab, CERN, FNAL...). These experimental programs (see Ref.~\cite{Hyd11} for a review), mostly using space-like virtual probes as in Deeply Virtual Compton Scattering (DVCS) or Deeply Virtual Meson Production (DVMP), will provide in a near future large but yet somewhat limited view-window on the hadronic structure. As a generalization of these programs, reactions with  final state time-like photons such as Time-like Compton Scattering (TCS)~\cite{Ber02} or Double Deeply Virtual Compton Scattering (DDVCS)~\cite{{Gui03},{Bel03}} appear as powerful new tools.

With advances of detector capabilities and of new experimental facilities that provide sufficiently high energy and 
high luminosity beams, lepton pair production becomes an attractive and fundamental process for nucleon and nuclear studies. 
At hadron facilities, the Drell-Yan process (DY)~\cite{Dre70} in the $p$-$p$ interaction is expected to provide essential 
tests of our understanding of the hadronic structure and enlightening data about the sea quarks in terms of Parton Distribution Functions (PDFs)~\cite{Gee06} and TMDs~\cite{Ado16}. At lepton facilities, for large enough virtualities and sufficiently small momentum transfer, the exclusive leptoproduction off the nucleon of a time-like photon decaying into a pair of leptons ($lp \to lp \gamma^* \hookrightarrow \ell^+\ell^-$), probes the internal quark and gluon structure of the nucleon expressed within the GPDs' framework~\cite{Bel03-1}. In addition, TCS provides a complementary and essential measurement of GPDs~\cite{Boe15} while at close-to-threshold kinematics heavy quarkonium (as $J/\Psi$) production probes the gluonic form factors of the nucleon~\cite{Fra02}. Alltogether, these reactions open an avenue for exploring the quark-gluon structure of hadrons at a new level. 

At sufficiently high energies, TCS corresponds to the production of a time-like virtual photon off a quark, with subsequent decay into a lepton pair $(\gamma N \rightarrow \ell^+ \ell^- N'$). The virtuality of the final photon provides the hard scale of relevance for the factorization of the reaction amplitude into a known hard part and an unknown soft part  parameterized in terms of GPDs, and corresponding to the non-perturbative long range interaction between quarks and gluons. At leading  order and leading twist, the amplitudes of DVCS ($lN \to lN\gamma$) and TCS are complex conjugate. This property allows to uniquely test universality of GPDs without any additional unknowns.

In contrast to DVCS and TCS reactions, where observables contain integrals of GPDs over the quark internal momentum fraction ($x$) and GPDs at skewness ($\xi$) related to specific momentum fraction ($x$=$\pm\xi$), the high virtuality of both the incoming and outgoing photons in DDVCS, at Born level,  allows to map out GPDs in a wide range of the $x \ne \xi$ phase space. The time-like virtuality of the final photon provides the additional kinematic lever arm to explore the out-of-diagonal phase space by varying the invariant mass of the decaying dilepton. The $(x,\xi)$ decoupled knowledge of GPDs is not only very valuable for constraining GPD models and fitting procedures, but is also of particular importance for nucleon imaging strictly defined at zero-skewness~\cite{Bur03}. 

DVMP accompany TCS and DDVCS not only as background channel but also as specific channel of interest for nucleon and nuclear structure studies. A particular case is the production and interaction of heavy quarkonia  with hadronic matter where the small spatial size of heavy quarkonia allows to describe their interactions with hadrons within controlled approximations. Heavy quarkonium production probes the local color (gluon) fields of the nucleon, and can reveal properties such as their response to momentum transfer, their spatial distribution, and their correlation with valence quarks. The dynamics that produces the relevant gluon fields in the nucleon changes considerably between high energies and the near-threshold region, creating a fascinating landscape calling for detailed experimental study. In the context of nuclear physics, studying dilepton decay of vector meson in nuclear matter is an important step for understanding the dynamics of nucleon-meson interactions and of the produced matter in heavy ion collisions, as the dilepton pair does not undergo strong interaction in the final state. 

Such measurements, which are just beginning to be within the reach of actual experimental capabilities, require a 
deeper theoretical understanding to provide efficient guidance for experiments and insure appropriate data interpretation. Several questions still remain open as, among others, the correct link between the TCS and DVCS processes, the access to 
dynamic properties of the nucleon, the complementarity between DDVCS and dispersion relation properties, the importance 
of vector meson production for nuclear structure studies etc... The International Workshop on Nucleon and Nuclear Structure through Dilepton Production was organized at the European Center for Theoretical Studies in Nuclear Physics and Related Areas (ECT$^{\star}$) of Trento during the period 24$^{\mathrm{th}}$-28$^{\mathrm{th}}$ October 2016. It gathered a community of theorists and experimentalists to discuss these issues, define, and optimize a strategy for different 
measurements towards the most significant impact. 

This article constitutes the proceedings of the ECT$^{\star}$ workshop, built from the contributions of each participant 
to deliver a road-map for future studies of the nuclear structure in the dilepton production channel. The next section 
reports the discussions about analytical properties of GPDs, of importance to establish the nature and the strength of 
the link between the different processes. The following sections discuss the different possibilities to experimentally 
access GPDs with the dilepton channel as well as novel features of Quantum ChromoDynamics (QCD) that may also be adressed. 
The section after is a brief review about the main approaches to modeling GPDs from first principles or extracting them from 
current existing data on deep exclusive processes. The following section describes the experimental projects currently under 
development at different facilities. Finally, on the basis of the preceeding information and of the discussions occuring during 
the workshop, a global strategy for dilepton experiments is elaborated.

%
%
\section{Analytical properties of GPDs}

Understanding of analytical properties of the scattering amplitudes of exclusive DVCS and of hard exclusive meson production (DVMP) within description based on the QCD factorisation is an important part of theoretical studies. This is due to the fact that knowledge of analytical properties of scattering amplitudes permits to clarify the relationship of subtraction constants in dispersion relation with the Polyakov-Weiss $D$-term~\cite{Polyakov:1999gs}. The knowledge of analytical properties permits also to understand the relation of generalized parton distributions (GPDs) with generalized distribution amplitudes appearing after analytic continuation of GPDs from the $s$-channel to the crossed $t$-channel.

Let us summarise the essential points of above issues \cite{Teryaev:2005uj}. The generic Compton form-factor 
$\cal{H}(\xi)$ of DVCS or longitudinal DVMP in Born approximaton has the form
\begin{equation}
\mathcal{H}(\xi) = \int\limits_{-1}^{1} dx  \;\frac{H(x,\xi)}{ x - \xi + i \epsilon}\;.
\label{CF}
\end{equation}
This expression is similar to a form of a dispersion relation in skewness $\xi$. However due to dependence on $\xi$ of 
the GPD $H(x,\xi)$ appearing in the numerator, Eq.~\ref{CF} is not a standard dispersion relation. The construction 
of dispersion relation for Compton form-factor  $\mathcal{H}(\xi)$ requires knowledge of its analytic properties in $\xi$. 
They are uncovered by noting that polynomiality property of the GPD $H(x,\xi)$ implies that 
\begin{equation}
\lim_{\xi \to \infty} \frac{1}{\xi^{n+1}} \int\limits_{-1}^{1}dx \,x^n H(x,\xi)
\end{equation}
is a constant which leads to the conclusion that for large $\xi \gg 1$ in the unphysical region the power series in $\xi$ of  the Compton form-factor 
\begin{equation} 
\mathcal{H}(\xi) = -  \sum\limits_{n=0}^\infty  \frac{1}{\xi^{n+1}}\int\limits_{-1}^{1} dx x^n H(x,\xi) 
\end{equation}
is convergent. This fact supplies the argument  that $\mathcal{H}(\xi)$ is an analytic function  for large $\xi$ with the cut for $\xi \in [-1,1]$. Thus the unsubtracted dispersion relation for $\mathcal{H}^{dr}(\xi)$ has the form
\begin{equation}
\mathcal{H}^{dr}(\xi) = \int\limits_{-1}^{1} dx  \;\frac{H(x,x)}{ x - \xi + i \epsilon}\;.
\label{CFDR}
\end{equation}
The difference between representations of Eq.~\ref{CF} and Eq.~\ref{CFDR} thus reads
\begin{eqnarray}
&&\Delta \mathcal{H}(\xi) = \mathcal{H}(\xi) - \mathcal{H}^{dr}(\xi) = 
 \\
&&\sum\limits_{n=0}^\infty \frac{1}{n!} \sum\limits_{k=0}^{n-1} \binom{n-1}{k} (-\xi)^{n-1-k}\frac{d^n}{d\xi^n} \int\limits_{-1}^1 dx x^k H(x,\xi) 
\nonumber
\label{diff}
\end{eqnarray}
and due to the polynomiality properties of GPD $H(x,\xi)$ the non-vanishing contribution to $\Delta \mathcal{H}(\xi)$  
comes only from the highest power $\xi^n$ of the moments of $H(x,\xi)$ which means that  $\Delta \mathcal{H}(\xi)$ is a constant independent of $\xi$. The origin  of such a constant can be attributed to the presence of a subtraction term in the  dispersion relation (Eq.~\ref{CFDR}) which resides entirely in the ERBL region. 

The relation between scattering amplitude for a space-like type process such as DVCS with a time-like process such as TCS can  be understood by examining their analytic properties. In LO approximation they are related by simple complex conjugation operation which leads to a change of $i\epsilon$ sign in propagators. In TCS at NLO approximation~\cite{PSW} the complex conjugation operation has to be supplemented by a term related to evolution of GPDs, for more details see Ref.~\cite{MPSzW}. This complication is related to the fact that whereas in DVCS the $s$-channel cut is responsible for all imaginary parts of hard amplitude, in the time-like case of TCS the imaginary part can be attributed either to a cut in the $s$-channel or the  cut related to the presence of a virtual photon with positive $Q^2$.

Moreover, partial cancellation between contributions coming from $s$- and $Q^2$- cuts can occur, which will result in a different $i\epsilon$ prescription than in the DVCS case (for the case of the hard coherent dijet production on hadrons this is discussed in Ref.~\cite{Braun:2002wu}).  

%
%
\subsection{$D$-term for various hard processes}
\vspace*{-4pt}
\subsubsection*{\hspace*{18pt} \small\em{O.V.~Teryaev}}
\vspace*{-4pt}
The $D$-term appears as a finite subtraction in the dispersion relations for DVCS and DVMP hard exclusive reactions at 
LO~\cite{Teryaev:2005uj} and NLO~\cite{Diehl:2007jb}.

For TCS there is a cancellation of cuts in $s$ and $Q^2$ in the ratio $Q^2/s$, similar to inclusive fragmentation, while the cuts in $Q^2$ in the ratio $Q^2/u$ acquire the form of cuts in $u$, explaining the opposite $i \varepsilon$ prescription found by other authors.

For DDVCS, where both cuts in $s$ and $Q^2$ meet, the $D$-term contributes also to the imaginary part of the amplitude, resulting in the additional $\xi$ independent term
\begin{equation}
\Im m [M] \sim H(\xi,\xi) + D(R) \, ,
\end{equation}
where 
\begin{equation}
R=(Q^2 - M^2)/(Q^2+M^2) \, .
\end{equation}

%
%
\subsection{The Origin of the $J=0$ Fixed Pole in QCD}
\vspace*{-4pt}
\subsubsection*{\hspace*{18pt} \small\em{S.J.~Brodsky}}
\vspace*{-4pt}
The instantaneous light-front (LF) $\gamma q \to \gamma q $ interaction $e^2_q \gamma^+ ({1\over k^+})^2 \gamma^+$ in the QCD LF Hamiltonian predicts a $J=0$ fixed pole $s$-independent contribution to virtual Compton scattering amplitudes $\gamma^* p \to \gamma^{*'} p'$ at all values of photon virtualities $Q^2$, $Q^{'2}$ and momentum transfer squared $t$, in analogy to the {\it seagull} interaction for Compton scattering on scalar charged fields~\cite{Brodsky:2008qu}. Since the resulting 
amplitude is real, it interferes maximally with the Bethe-Heitler amplitudes~\cite{Brodsky:1972vv}.

%
%
\hyphenation{pros-pects}
\subsection{Dual parameterisation of GPDs and the Mel\-lin-Bar\-nes transform approach}
\vspace*{-4pt}
\subsubsection*{\hspace*{18pt} \small\em{K.M.~Semenov-Tian-Shansky}}
\vspace*{-4pt}
Present day strategy for extracting GPDs from the data relies on employing of phenomenologically motivated GPD representations which allow to implement the fundamental requirements such as polynomiality, hermiticity, positivity {\it etc.} following from the underlying field theory. A class of convenient GPD representations is based on the expansion over the conformal partial wave  basis, which ensures the diagonalization of the leading order evolution operator. The dual parameterisation of GPDs~\cite{Polyakov:2002wz} and the Mellin-Barnes integral approach~\cite{Mueller:2005ed} represent the two frameworks for handling the double partial wave expansion of GPDs in the conformal partial waves and in the cross-channel ${\rm SO}(3)$ partial waves. In  Ref.~\cite{Muller:2014wxa} we demonstrated the complete equivalence of these two independently developed representations. 

This finding provided additional insight into the GPD properties and, in particular, allowed  to clarify the relation between the $D$-term form factor and the $J=0$ fixed pole contribution into the DVCS amplitude. The $J=0$  
fixed pole universality hypothesis formulated in Ref.~\cite{Brodsky:2008qu} fixes the value of this contribution in 
terms of the analytically regularized inverse Mellin moment of $t$-dependent PDF
\begin{equation}
a_{J=0}^{\rm f.p.}(t)=-2 \int^{1}_{(0)} \frac{dx}{x}\, q^{(+)}(x,t)
\end{equation}
where $(0)$ denotes the analytic regularization prescription. In Ref.~\cite{Muller:2015vha}
we argued that this hypothesis lacks rigorous proof. 

However, the detailed experimental measurements of DVCS both in space-like and time-like regimes, and prospects to study experimentally the doubly deeply virtual Compton scattering open the perspective for the direct experimental check of the 
$J=0$ fixed pole universality hypothesis which is very important for our understanding of QCD by means of the sum rule
\begin{equation}
a_{J=0}^{\rm f.p.}(t) = 4 D(t|\vartheta) - 2 \int^{1}_{(0)} \frac{d x}{x}\,  H^{(+)}(x,\vartheta x,t) 
\label{a_{J=0}^{f.p.}}
\end{equation}
where ${\cal D}(t|\vartheta)$ stands for the $D$-term form factor, $\vartheta = (Q^2-Q^{'2}) / (Q^2+Q^{'2})$ is the photon  asymmetry parameter, and $H^{(+)}(x,\vartheta x,t)$ is the GPD which is on the cross-over line ($x=\eta$)
for the DVCS case ($\vartheta=1$).  

%
%

\section{Access to GPDs}

%
%
\subsection{Time-like Compton Scattering}
\vspace*{-4pt}
\subsubsection*{\hspace*{18pt} \small\em{J.~Wagner}}
\begin{figure}[!t]
\begin{center}
\resizebox{0.230\textwidth}{!}{%
  \includegraphics{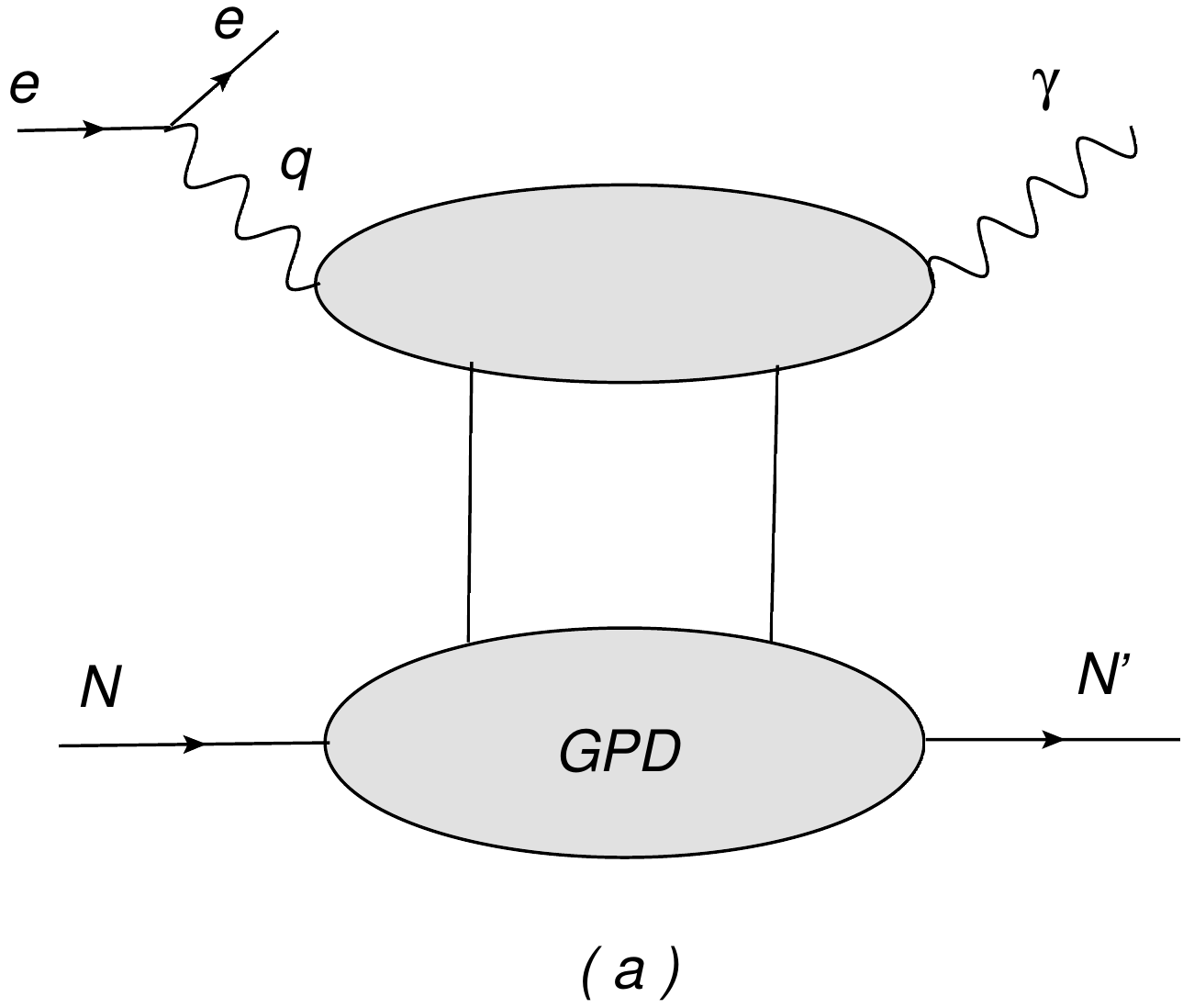}
 }
 \hspace*{0.3cm}
 \vspace*{-1.0cm}
  \resizebox{0.230\textwidth}{!}{%
  \includegraphics{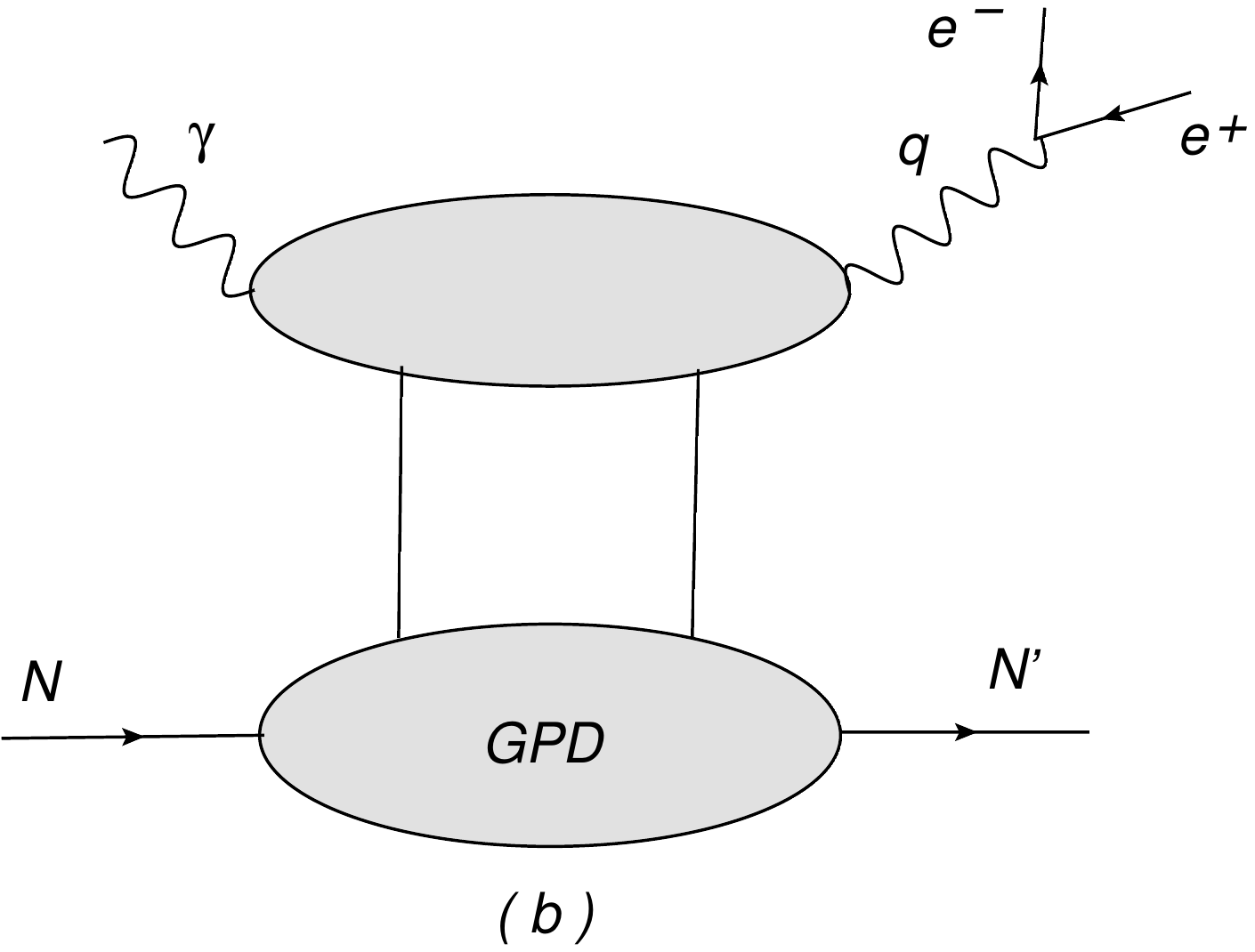}
  }
\end{center}
\vspace*{.5cm}
\caption{The DVCS (a) and TCS (b) processes are linked by time reversal and analyticity. They factorize in hard coefficients (upper blob) and generalized parton distributions (lower blob).}
\label{fig1-diag}
\end{figure}

\vspace*{-4pt}
The exclusive photo-production of a lepton pair through (TCS)~\cite{{Ber02},{Boe15}} shares many features with DVCS (Fig.~\ref{fig1-diag}). It may serve as a new source of knowledge about generalized parton distributions (GPDs), but also it is the cleanest channel in which we may test the universality of the GPD description of the hard exclusive processes. Contrary to the amplitude of DVMP, which contains also the non-perturbative distribution amplitudes, in the case of TCS and DVCS the only unperturbative part is GPD, and the only difference between them comes from the structure of the hard scattering coefficient functions. Those coefficient functions are known at the NLO accuracy in the strong coupling constant, and the relation between them reflects the differences in analytical structure of the amplitudes with large space-like and large time-like scales~\cite{MPSzW}.

As in the case of DVCS, a purely electromagnetic competing mechanism, the Bethe-Heitler (BH) mechanism contributes at the amplitude level to the same final state as TCS. This BH process cross section overdominates the TCS, and one has to restrict to more differential observables sensitive to the interference term. One of such observables $R$, defined in~\cite{{Ber02},{Boe15}}, utilises the angular dependence of the final state leptons and is sensitive to the real part of the Compton form factor $\mathcal{H}$, which is relatively difficult to determine for example in DVCS. The other observable, circular photon asymmetry which is sensitive to the imaginary part of $\mathcal{H}$, is measurable in the case of polarized electron beam such as CEBAF. Also the less known $\widetilde{\mathcal{H}}$ can be extracted from the TCS measurement, making use of the linear polarization asymmetry \cite{GPW} of real photons.  

Phenomenological studies~\cite{MPSSzW} demonstrated the important role of the NLO corrections for TCS observables. Size of those corrections, especially for the small values of the genaralized Bjorken variable, suggest a need for a high energy resummation~\cite{IPSzW}.
\unitlength 1mm

%
%
\hyphenation{Szy-ma-now-ski}
\subsection{Selected topics in Drell-Yan theory: from twist 3 SSA to exclusive limit}
\vspace*{-4pt}
\subsubsection*{\hspace*{18pt} \small\em{O.V.~Teryaev}}
\vspace*{-4pt}
The proper treatment of electromagnetic gauge invariance related to contour gauge for gluonic field results in extra factor 2 for DY single spin asymmetry and generation of gluonic poles by the physical components of gluon fields only~\cite{Anikin:2010wz}. The factor 2 may in principle be checked in the transverse polarized DY at COMPASS, and later, at J-Parc and NICA.

Exclusive DY process, besides the contribution of pion DA and GPD, suggested and explored by Pire, Szymanowski, Goloskokov, Kroll and others, involves also the contribution of competing mechanism of pion-proton DY scattering expressed in terms of two GPDs~\cite{Teryaev:2005uj}. The infrared stable part of this contribution may interfere with pure electromagnetic 
process~\cite{Pivovarov:2015vya}. The interference term, while integrating to zero when the average over dilepton angles is performed, leads to the charge asymmetry of dileptons of the several percent order.

%
%
\subsection{Probing GPDs through exclusive photo-production of $\gamma \rho$ pairs with large invariant mass}
\vspace*{-7pt}
\subsubsection*{\hspace*{18pt} \small\em{R.~Boussarie}}
\vspace*{-18pt}
\begin{figure}[!h]
\centerline{
 \includegraphics[width=0.3\textwidth]{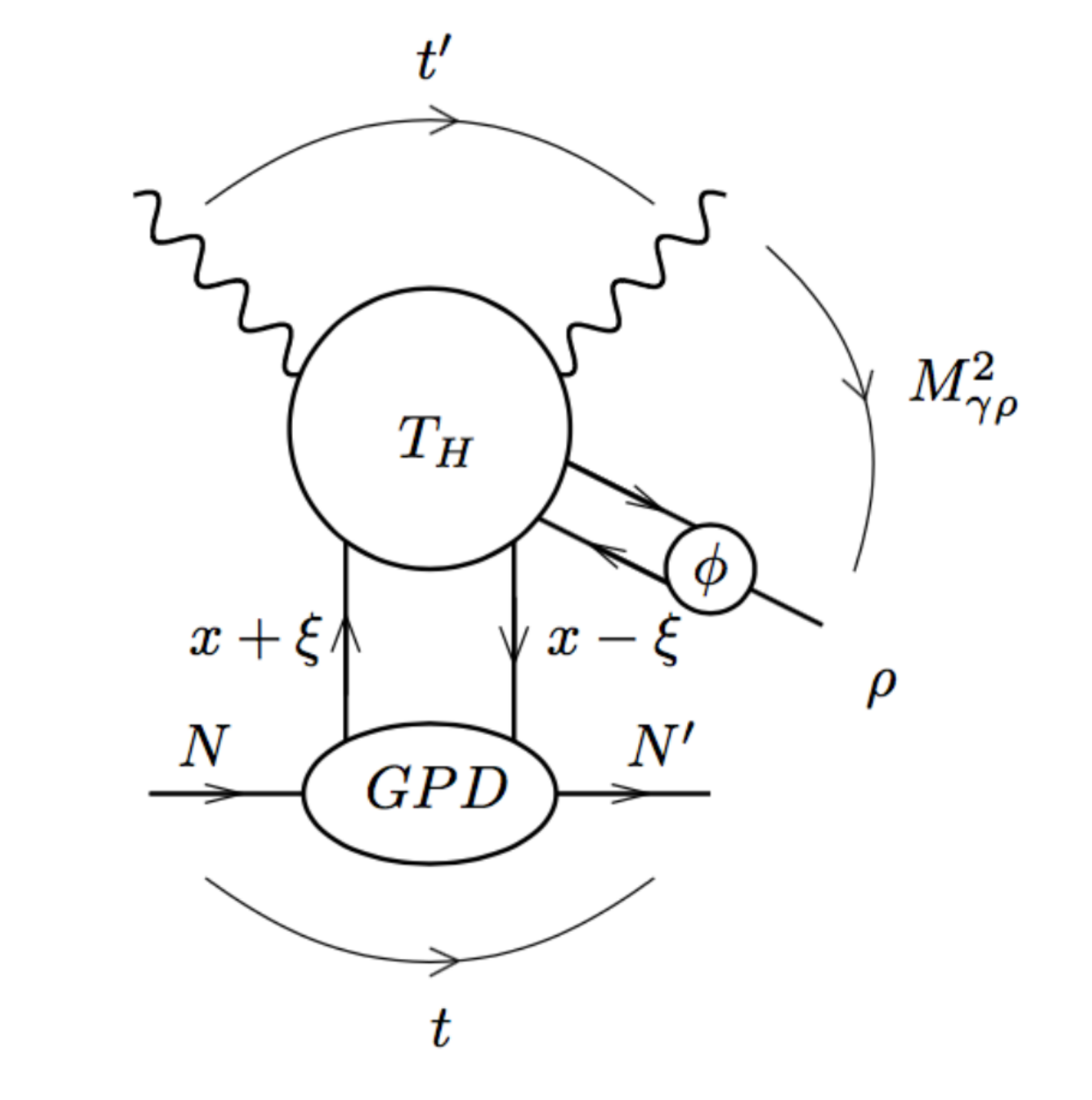}
}
\vspace*{-5pt}
\caption{Factorization of the amplitude for $\gamma + N \rightarrow \gamma + \rho +N'$ at large $M_{\gamma\rho}^2$.}
\label{Fig:feyndiag}
\end{figure}
In Ref.~\cite{Boussarie:2016qop} a feasibility study for the photo-production of a $\gamma\rho$ pair with a large invariant mass $(\gamma+N\rightarrow \gamma+\rho+N^\prime)$ at CLAS12 was performed. Such a process, for which one can apply the collinear factorization formalism (Fig.~\ref{Fig:feyndiag}) of QCD, gives a strongly motivated complement to the emblematic DVCS and DVMP reaction: it allows access in principle to the charge conjugation odd part of all GPDs, with a strong dominance of some helicity amplitudes (namely the production of a longitudinal $\rho$-meson, which involves $H$ and $\widetilde{H}$ GPDs). The high statistics which are predicted ($6.8\,10^6\,\rho_L$ and $7.5\,10^3\,\rho_T$ for 100 days of run with a proton target) will allow a consistent check of the universality of chiral even GPDs in a valence quark dominated process, using quasi-real photon beams to access a hard exclusive process similarly to TCS. More refined theoretical predictions in the future could allow one to get access to the elusive chiral-odd GPDs.\\
Such a feasability study may be extended to other experiments at JLab 12~GeV and COMPASS~\cite{Abb07}, and it will allow for predictions for the EIC in the future.

%
%
\subsection{$J/\psi$ photo-production on nuclei}
\vspace*{-4pt}
\subsubsection*{\hspace*{18pt} \small\em{V.~Guzey}}
\begin{figure}[t]
\begin{center}
\resizebox{0.45\textwidth}{!}{\includegraphics{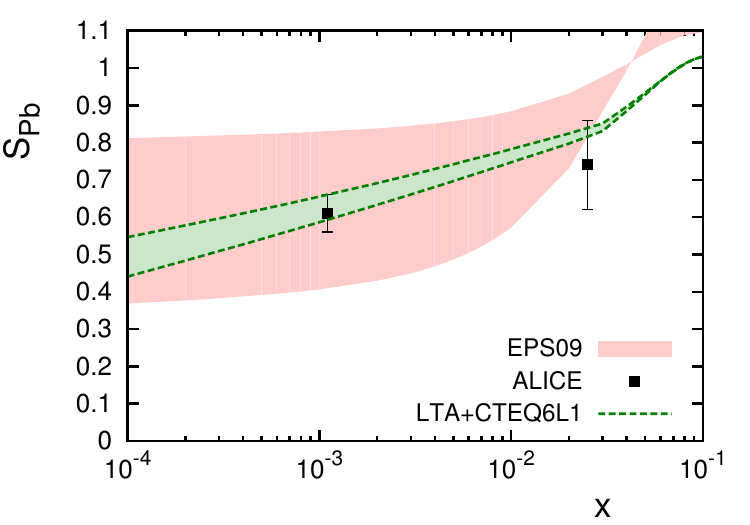}}
\caption{The nuclear suppression factor of $S_{Pb}$ (Eq.~\ref{SPb}) with experimental values extracted from ALICE data, theoretical predictions of the leading twist nuclear shadowing model, and the EPS09 fit to nuclear PDFs.}
\label{fig:S_pb208}
\end{center}
\end{figure}

\vspace*{-4pt}
In ultraperipheral collisions (UPCs), relativistic ions interact at large impact parameters, which suppresses the strong interaction and leads to the interaction via an exchange of quasi-real photons. Thus, UPCs at the Large Hadron Collider (LHC) allow one to study photon--proton and photon--nucleus interactions at unprecedentedly high energies~\cite{Baltz:2007kq}. With nuclear targets, a key process is coherent photo-production of $J/\psi$ vector mesons on nuclei, which probes the nuclear gluon distribution $g_A(x,\mu^2)$ at small values of the momentum fraction $x$, where  $g_A(x,\mu^2)$ is unconstrained by the existing data. To the leading orders in the strong coupling constant and the non-relati\-vis\-tic expansion for the charmonium distribution amplitude, one can show that the factor describing the nuclear suppression of the $\gamma A \to J/\psi A$ cross section 
\begin{equation}
S_{Pb}=(\sigma_{\gamma A \to J/\psi A}/\sigma^{\rm IA}_{\gamma A \to J/\psi A})^{1/2} \label{SPb}
\end{equation} 
is directly proportional to the ratio of the nucleus-to-nucleon gluon distributions~\cite{Guzey:2013xba,Guzey:2013qza} 
\begin{equation}
R_g=g_A(x,\mu^2)/[A g_N(x,\mu^2)] \, . 
\end{equation} 
Figure~\ref{fig:S_pb208} shows the good agreement between the values of $S_{Pb}$ extracted from the ALICE data on 
$J/\psi$ photo-production in Pb-Pb UPCs at $\sqrt{s_{NN}}=2.76$ TeV and the predictions of the leading twist nuclear shadowing model~\cite{Frankfurt:2011cs} (LTA+CTEQ6L1) and the EPS09 fit to nuclear PDFs. It gives first direct evidence of large nuclear gluon shadowing at $x \approx 0.001$ and $\mu^2 \approx 3$ GeV$^2$. This can be further tested in run 2 at the LHC by measuring UPCs accompanied by forward neutron emission, and the momentum transfer distribution of the $A A \to J/\psi AA$ process which is predicted to be shifted toward smaller values of $|t|$ by the gluon nuclear shadowing~\cite{Guzey:2016qwo}.

%
%
\subsection{Accessing the real part of the amplitude of forward Compton scattering off the proton}
\vspace*{-4pt}
\subsubsection*{\hspace*{18pt} \small\em{O.~Gryniuk \& M.~Vangerhaeghen}}
\vspace*{-4pt}
We describe an approach of a direct experimental assessment of the real part ($\Re e[T]$) of the spin-averaged forward Compton scattering amplitude off the proton through dilepton photo-production process. The method discussed was already applied for such a measurement at DESY in 1973~\cite{Alvensleben}, which so far is the only existing datapoint, at photon beam energy 2.2~GeV. We propose to widen the accessed energy range and improve on the precision of such a measurement by engaging existing facilities such as Jefferson Lab. 

\begin{figure}[!t]
\begin{center}
  \vspace*{-7pt}
  \includegraphics[width=0.385\textwidth]{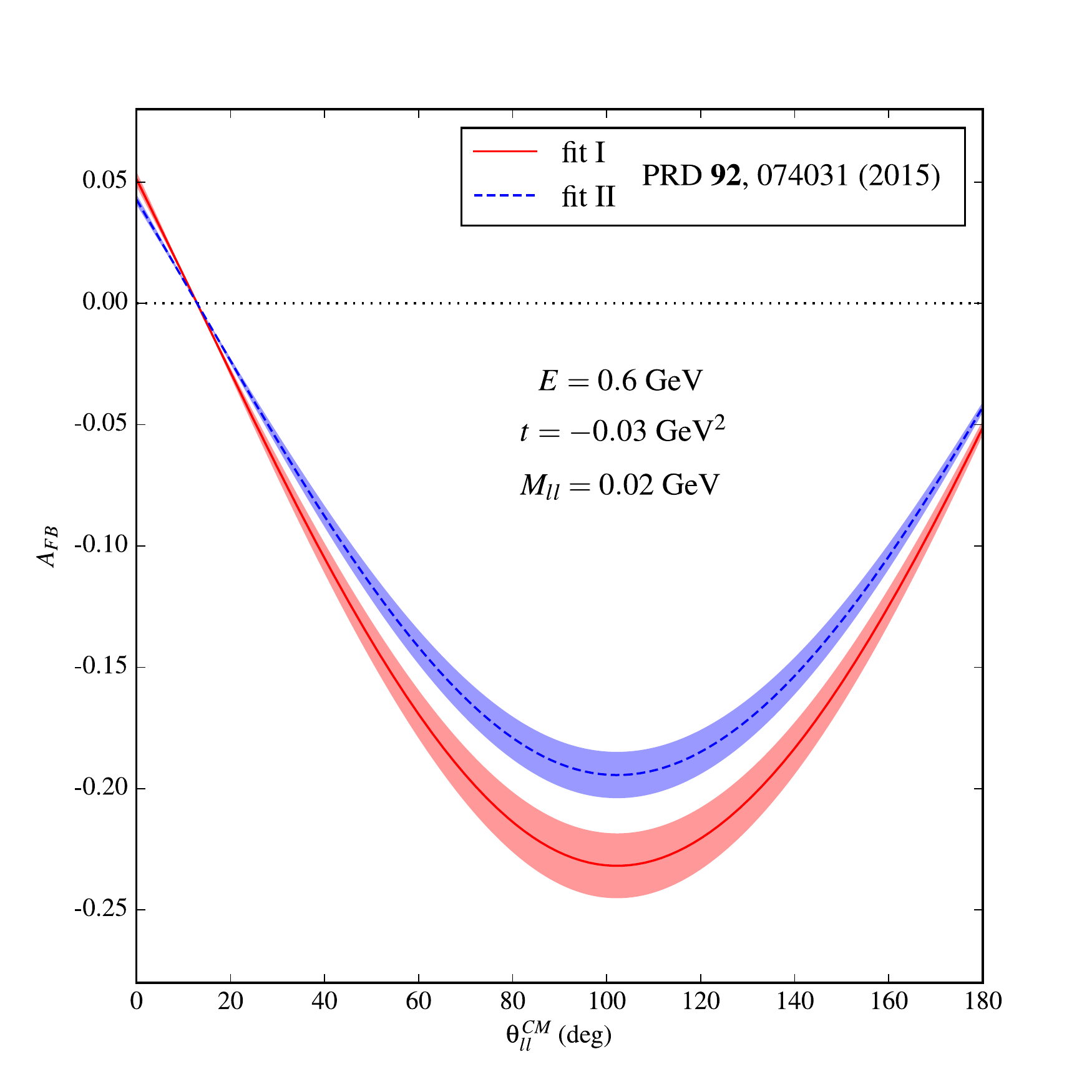}
  \includegraphics[width=0.385\textwidth]{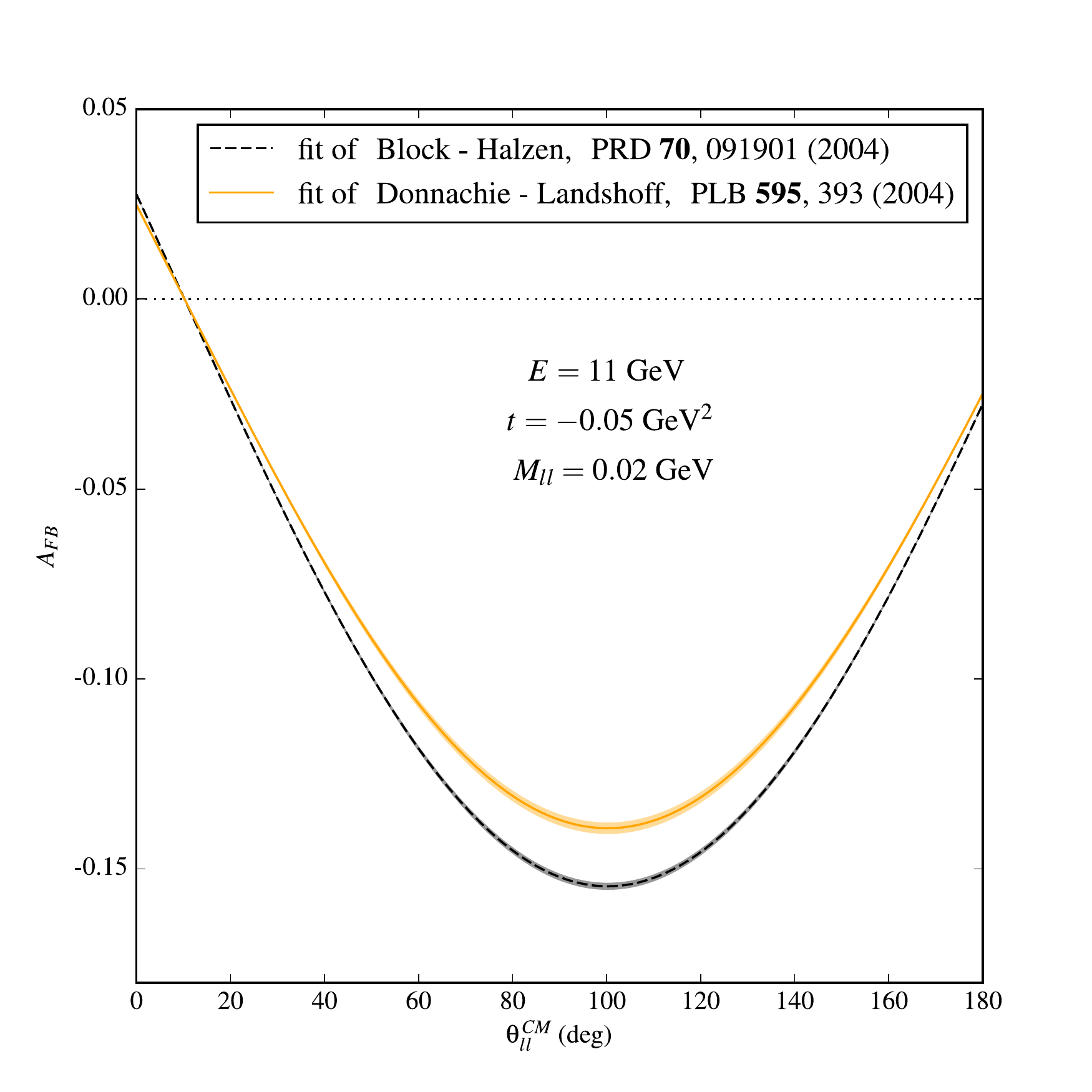}
 \caption{Forward-backward asymmetry for the $\gamma \, p \to e^- e^+ \, p$ process as a function of the lepton polar angle in lepton pair CM frame for different fits of $\sigma_\mathrm{tot}$ and for different kinematics. The bands represent the propagated uncertainty estimate.
}
\label{fig:AFB}
\end{center}
\end{figure}
Based on the fits of the total photo-production cross section, by using a once-subtracted dispersion relation, we construct $\Re e[T]$. We discuss the impact of various fits for low-~\cite{Gryniuk:2015} and high-energy~\cite{Block-Halzen:2004,Donnachie-Landshoff:2004} data and argue the need for a new direct amplitude assessment. We present an estimate of the forward-backward asymmetry to the $\gamma p \to e^- e^+ p$ process for low values of momentum transfer $-t$ and lepton pair mass $M_{ll}$, which results from interchanging the leptons in the interference between the time-like Compton and the Bethe-Heitler mechanisms. We show that this asymmetry can reach values around -20\% for reasonable kinematics (Fig.~\ref{fig:AFB}) providing a sensitive observable to directly access $\Re e[T]$.

%
%
\subsection{Novel features of QCD revealed in dilepton pro\-duc\-tion and DVCS}
\vspace*{-4pt}
\subsubsection*{\hspace*{18pt} \small\em{S.J.~Brodsky}}
\vspace*{-4pt}
{\small\bf\em Positronium-proton scattering and doubly-space-like virtual Compton scattering}

\begin{figure}[!ht]
\begin{center}
\resizebox{0.41\textwidth}{!}{\includegraphics{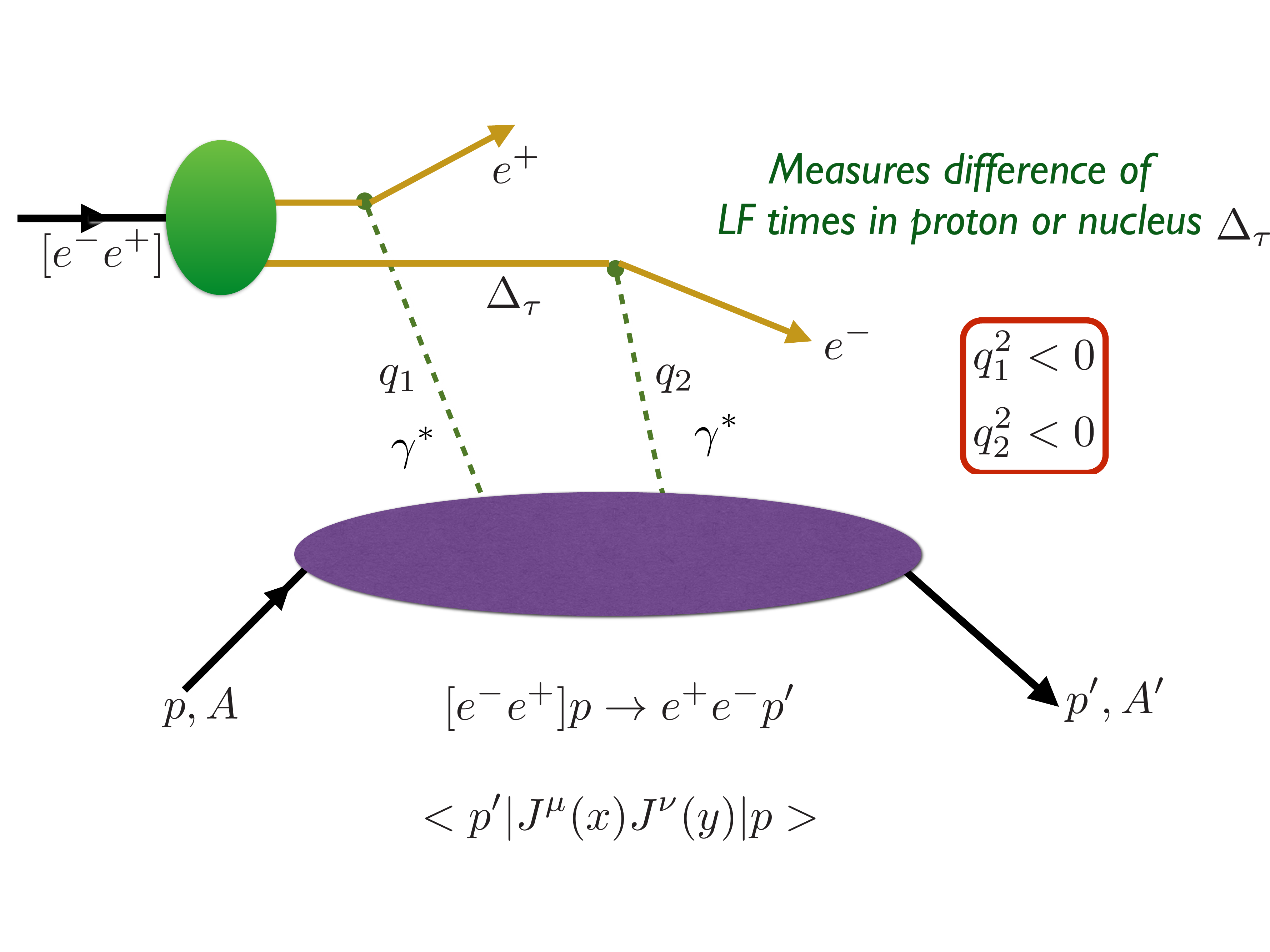}}
\end{center}
\vspace*{-15pt}
\caption{Measurement of doubly-space-like virtual Compton scattering using relativistic positronium beams.}
\label{fig1-bro}
\end{figure}

Doubly virtual Compton scattering on a proton (or nucleus) can be measured for two space-like photons with minimal, tunable, skewness $\xi$ using positronium-proton scattering $[e^+ e^-] p \to e^+ e^- p'$ (Fig.~\ref{fig1-bro}). In that process, the two lepton-quark interactions occur at separate LF times. The imaginary part of the $\gamma^* p \to \gamma^* p$ forward Compton amplitude gives the inelastic lepton proton cross section. The real part of the amplitude contains the $J=0$ fixed pole from the LF instantaneous quark exchange interaction. The same double-space-like amplitude contributes to the two-photon exchange contribution to the muonic hydrogen Lamb Shift.

One can also measure double deep inelastic scattering and elastic $[e^+ e^-]$-$p$ scattering. The inelastic scattering amplitude $[e^+ e^-] p \to e^- e^+ X$ measures two-parton deep inelastic lepton-proton scattering. Relativistic positronium beams can be created using Bethe-Heitler $e^+ e^-$ pair production just below the $e^+ e^-$ continuum threshold. An analogous  process will create the true muonium $[\mu^+\mu^-]$ atom~\cite{Brodsky:2009gx,Banburski:2012tk}.

{\small\bf\em Light-Front Theory}

\vspace*{4pt}
The boost-invariant (LFWF) Light-Front Wave Functions of hadrons $\psi(x_i, \vec {k_{\perp i}}, \lambda_i)$, that are the  eigensolutions of the QCD Light-Front Hamiltonian~\cite{Brodsky:1997de}, are the basis for computing~\cite{Brodsky:2000xy} the fundamental distributions $E(x, \xi, Q^2), H(x,\xi, Q^2)$  {\it etc.} that underly DVCS. LF quantization at fixed LF time $\tau=t+z/c$ provides a rigorous formulation of hadron physics, independent of the observer's frame. There is no Lorentz contraction of LFWFs \cite{Penrose:1959vz,Terrell:1959zz}, and the LF vacuum is trivial up to zero modes. There exists no expression analogous to the Drell-Yan-West overlap formula for the current matrix elements or the form factors of hadrons using ordinary fixed-time quantization, because of the need for dynamical boosts and the necessity to include acausal, vacuum-induced currents. 

{\small\bf\em Breakdown of factorization theorems from initial- and final-state interactions}

\vspace*{4pt}
The handbag approximation to DVCS will be broken by the same leading twist rescattering interactions that produce the Sivers effect in deep inelastic lepton-proton scattering~\cite{Brodsky:2002cx} and also diffractive deep inelastic lepton scattering $\gamma^* p \to X p$~\cite{Brodsky:2002ue}.  The double initial scattering Boer-Mulders effect destroys conventional factorization theorems such as the Lam Tung relation for the angular distribution of DY lepton pair production~\cite{Boer:2002ju}.

%
%
\subsection{Double DVCS}
\vspace*{-4pt}
\subsubsection*{\hspace*{18pt} \small\em{M.~Guidal}}
\vspace*{-4pt}
The Double Deep Virtual Compton scattering process corresponds to the exclusive electro-production of a lepton pair on the nucleon, i.e. the $e \: N \to \:\: e' N' \gamma' \hookrightarrow \ell^- \: \ell^+$ reaction. When the virtuality of the incoming ($Q^2$) or outgoing ($Q^{'2}$) virtual photon is large enough, it allows one to access the GPDs of the nucleon in a unique way. Namely, at QCD leading-twist and at leading-order, the three variables upon which the GPDs depend, i.e. $(x,\xi,t)$, can be indepedently accessed, which is not the case with the DVCS and TCS processes where the $x$-dependence is in general integrated over.

The imaginary part of the scattering amplitude for DDVCS at leading order, $i \Im m \left[ {\cal T}^{DDVCS}(2\xi'-\xi, 
\xi, t) \right]$, is a function of three variables $\xi'= x_B/(2-x_B)$ (where $x_B$ is Bjorken variable), $\xi = \xi' (Q^2 +Q^{'2})/{Q^2}$, and a square of momentum transfer $t$. Being proportional to a quark GPD 
\begin{equation}
H^{q(+)}=H^q(x,\xi,t) - H^q(-x,\xi,t)
\end{equation} 
it reads
\begin{eqnarray}
 &&i \Im m \left[ {\cal T}^{DDVCS}(2\xi'-\xi,\xi,t) \right] = \label{sign}
\\
&&\hspace*{2cm}-i \pi H^{q(+)}( \xi \frac{Q^2-Q^{'2}}{Q^2 + Q^{'2}},\xi,t)\,.
\nonumber
\end{eqnarray}
From Eq.~\ref{sign} and the antisymmetry property of the GPD $H^{q(+)}$ with respect to $x$, it follows that the imaginary part in the euclidean DVCS regime with $Q^2 > Q^{'2} $ and in the time-like Compton scattering regime with $Q^{'2} > Q^{2}$  have opposite signs.

\begin{figure}[!t]
\vspace*{-4.45cm}
\begin{center}
\resizebox{0.480\textwidth}{!}{%
\includegraphics{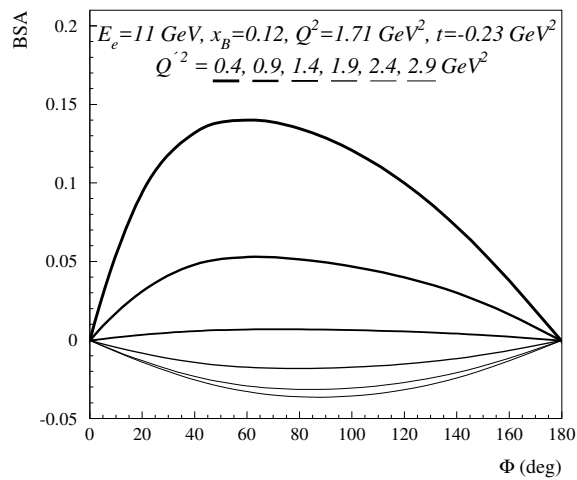}
}
\caption{Beam spin asymmetry (BSA) for the BH+DDVCS process as a function of the azimuthal angle between the electron scattering plane and the hadronic plane for typical JLab kinematics and different $Q'^2$ values (the angles of the 
decay lepton pair are integrated over). The thickness of the curves decreases as $Q'^2$ increases.}
\label{fig:bsa_phi}
\end{center}
\end{figure}

Therefore DDVCS process provides a powerful way to test the GPD formalism. It is indeed predicted that the beam spin asymmetry should have an opposite sign in the $Q^2>Q'^2$ and $Q^2<Q'^2$ regions (Fig.~\ref{fig:bsa_phi}), due to the change of sign of the imaginary part of the DDVCS amplitude which interferes with the real part of the Bethe-Heitler (BH)
amplitude. The interpretation of the process is the most straightforward when the final leptons are muons, which avoids complex antisymmetrization issues.

%
%
\subsection{Transversity GPDs}
\vspace*{-4pt}
\subsubsection*{\hspace*{18pt} \small\em{L.~Szymanowski}}
\vspace*{-4pt}
It is widely accepted that GPDs give access to the internal structure of hadrons in a much more detailed way than PDFs 
measured in inclusive processes, since they allow a 3-dimensional analysis. The appealing feature of description based 
on GPDs is that one can probe spin related quantities such as the elusive chiral odd helicity or transversity GPDs also 
in reactions on an unpolarized nucleon. Most of attempts which has been dedicated to studies of transversity GPDs within 
the domain of virtual photon mediated processes~\cite{{transGPDacc},{Coll00},{Ivan02},{Enbe06},{Beiy10}} seem to indicate that such studies are likely to be out of abilities of present accelerators and only future electron-ion colliders or the study of charmed meson production in neutrino processes~\cite{{neutrino},{PSW17}} may help. The promising exception in this respect is the proposal~\cite{{pion},{Golo10},{Pire05},{Lans12},{Pire13}} arguing that experimental access to transversity GPDs already at JLab can be achieved in pion electro-production due to their interplay with twist-3 distribution amplitudes of the pseudoscalar mesons. 

%
%
\section{Modeling and determination of GPDs}

With the now existing amount of GPDs data gathered over the last few years fitting methods are needed in order to achieve the goal of extracting either GPDs or GPD properties and constrained their modelling. The double-distribution (DD) formalism is the most popular model in use. Several methods are applied but the knowledge of the actual GPDs is still limited, especially concerning the $D$-term. The main features of this effort are reviewed hereafter. 

%
%
\subsection{Finite-$t$ and target mass corrections}
\vspace*{-4pt}
\subsubsection*{\hspace*{18pt} \small\em{V.~Braun}}
\vspace*{-4pt}
In the last years, a systematic approach to calculate kinematic higher-twist corrections $t/Q^2$ and $M^2/Q^2$ to 
hard exclusive processes in off-forward kinematics has been developed~\cite{Braun:2011zr,Braun:2011dg,Braun:2012bg,Braun:2012hq,Braun:2014sta}. A detailed analysis of the potential 
impact of such corrections on various DVCS observables are presented in~\cite{Braun:2014sta}. The same technique~\cite{Braun:2011zr,Braun:2011dg,Braun:2012bg} can be applied to the time-like case and also to DDVCS but detailed
expressions still need to be worked out. The main features of the DVCS results are the following. First and most importantly, QCD factorization holds including kinematic twist-four corrections, although it is generally violated at this accuracy level. Electromagnetic gauge and Lorentz (translation) invariance is restored up to twist-five effects.  Target mass corrections are mostly absorbed in the dependence on the minimal kinematically allowed value of the momentum transfer ($t_{\rm min}$) and are in general rather small. The correct behavior of all helicity amplitudes at threshold $t\to t_{\rm min}$ is reproduced.  Finite-$t$ corrections are, on the contrary, rather large and must be taken into account in all studies aiming to uncover the three-dimensional nucleon structure. At a qualitative level, the main effect of kinematic corrections in fixed target experiments seems to indicate a preferred role of the frame of reference where the two photon momenta are used to define the  longitudinal plane. In collider experiments, a $t$-dependent redefinition of the relation between Bjorken $x$ variable and the skewness parameter $\xi$ is another large effect. A comparison with the recent data by the Jefferson Lab Hall A  Collaboration can be found in~\cite{Defurne:2015kxq}. Based on the existing studies we estimate that for most observables the region $t/Q^2 < 0.25$ is safe for using QCD factorization techniques involving generalized parton distributions. The theory of kinematic corrections can be improved in future in several aspects: the resummation of $(t/Q^2)^k$ corrections to all twists but still at the leading order (LO) in strong coupling is feasible, whereas the generalization to NLO is complicated and a more distant task. 

%
%
\subsection{The double-distribution pa\-ra\-me\-te\-ri\-sa\-tion of GPDs}
\label{sec:2}
\vspace*{-4pt}
\subsubsection*{\hspace*{18pt} \small\em{P.~Kroll}}
\begin{figure}[!t]
\begin{center}
\resizebox{0.235\textwidth}{!}{%
  \includegraphics{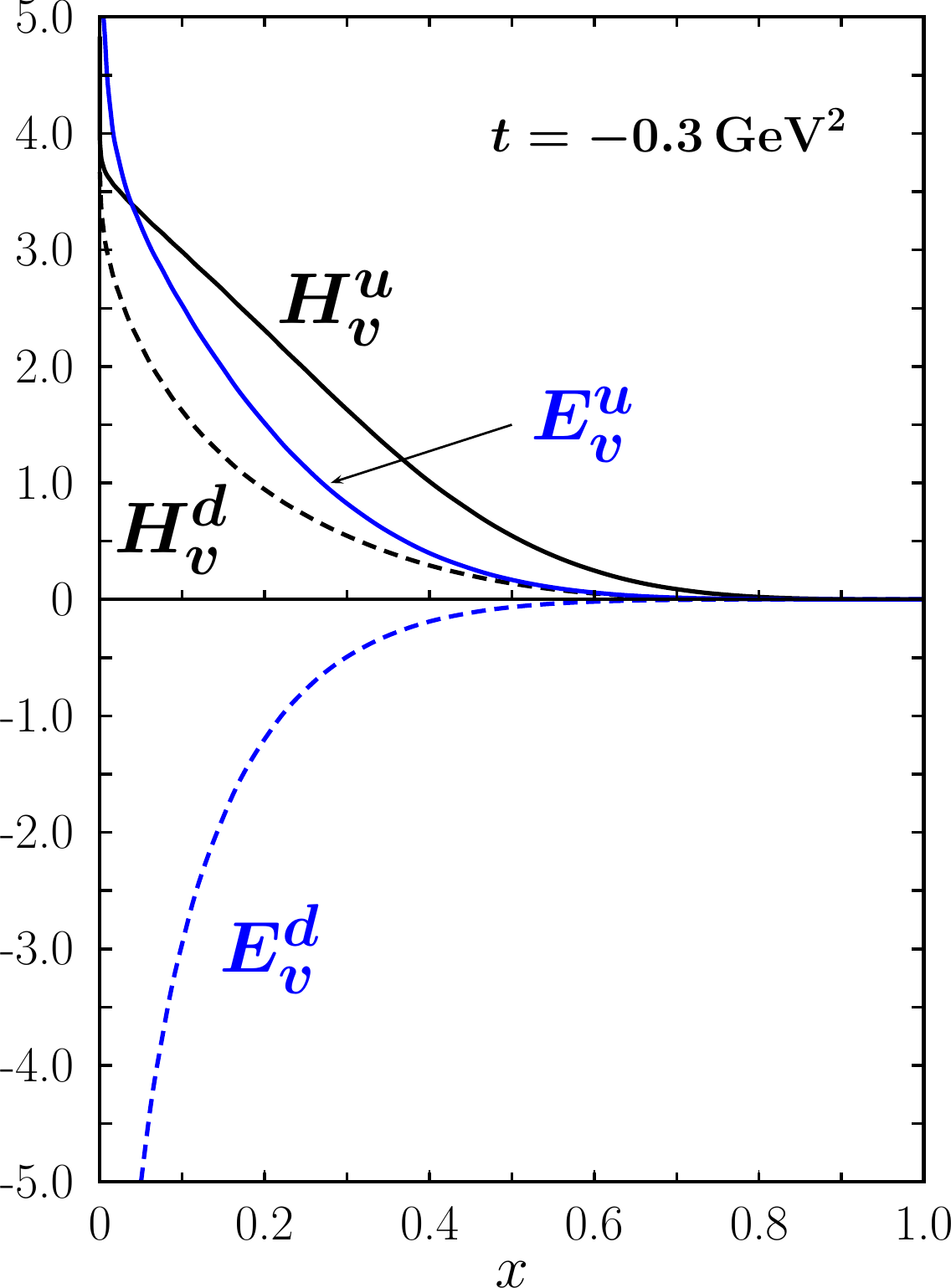}
  }
  \resizebox{0.235\textwidth}{!}{%
  \includegraphics{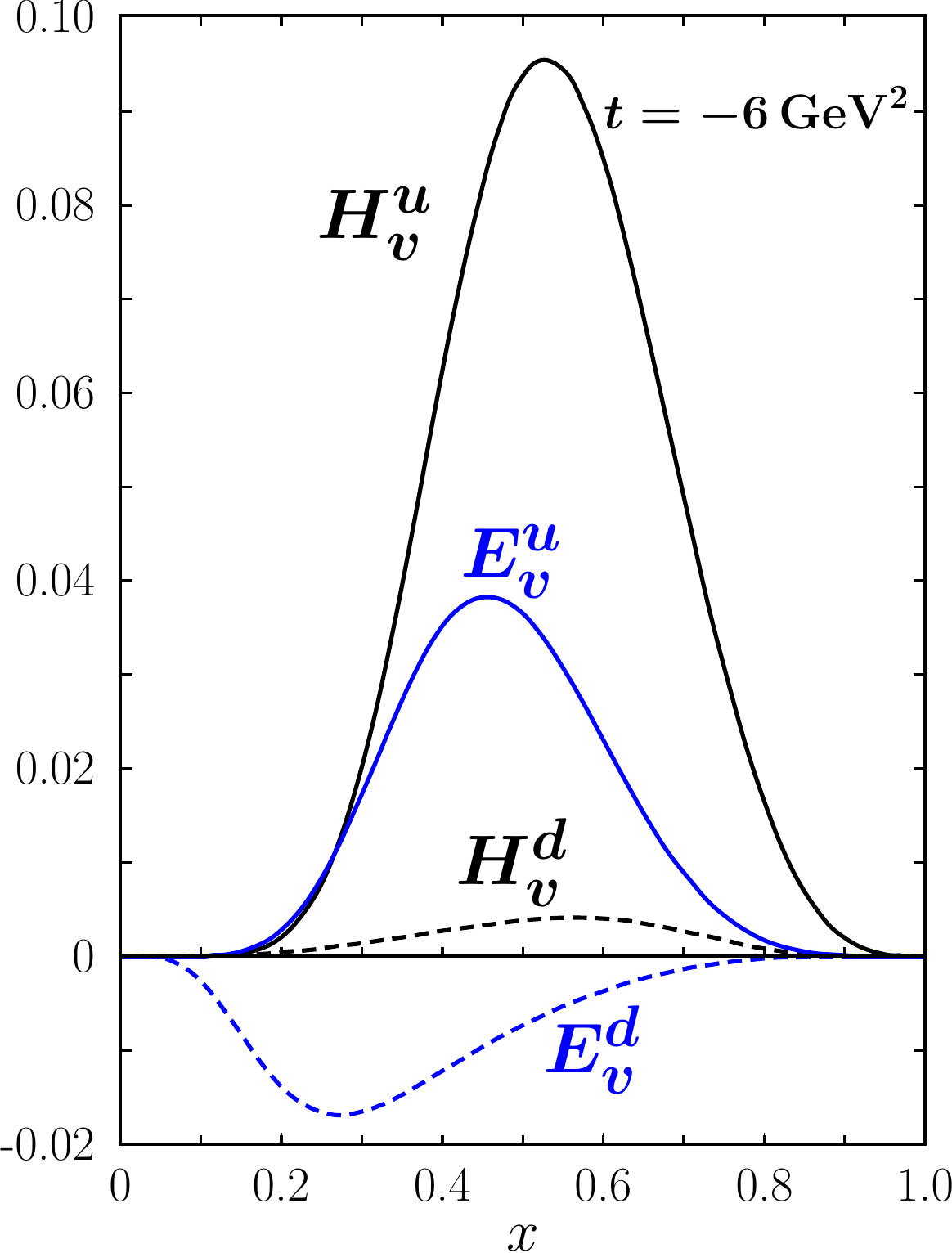}
  }
\caption{The valence-quark GPDs at two different values of $t$~\cite{Diehl-KrollarXiv:1302.4604}.\label{fig:Kroll}}  
\end{center}
\end{figure}

\vspace*{-4pt}
The double-distribution representation of generalized parton distributions is frequently used to pa\-ra\-mete\-rise the GPDs. According to Radyushkin the DD is assumed to be a product of the zero-skewness GPD and a weight function that generates the skewness dependence. Several ansaetze for the zero-skewness GPDs have been discussed: the $x - t$ factorized ansatz which is in severe conflict with large-$x$ properties of the GPDs; the familiar Regge-like parametrization $q(x)\exp{[f_q(x)t]}$ which is suitable to fit data on DVCS and DVMP at small $-t$. A more complicated $f_q$ profile function is needed for large $-t$ data in order to match all known properties of the GPDs. An important feature of the GPDs obtained from the latter two ansaetze is the strong $x-t$ correlation (Fig.~\ref{fig:Kroll}).   

The gluon GPD has also been discussed. Since vector-meson electro-production at small skewness is diffractive, the small 
$x$-behavior ($xg(x)\sim x^{-\delta_g(Q^2)}$) of the gluon den\-si\-ty is directly related to the energy dependence of the integrated cross section ($\sigma_L\sim W^{4\delta_g(Q^2)}$) at fixed $Q^2$. As a matter of fact the power $\delta_g(Q^2)$ 
is universal, i.e. independent on the vector meson. 

%
%
\subsection{The extraction of GPDs from DVMP}
\vspace*{-4pt}
\subsubsection*{\hspace*{18pt} \small\em{P.~Kroll}}
\vspace*{-4pt}
It is an experimental fact that DVMP is subject to strong power corrections for photon virtualities between about 2 and 20 $\mbox{GeV}^2$. The transverse amplitudes are not small, even dominant for pion and $\omega$ production, in that region of $Q^2$. There are even power corrections to the longitudinal amplitudes. For instance, the pion pole contribution to $\pi^+$ production is underestimated if evaluated from $\widetilde{E}$. We suggested that the transverse size of the meson is responsible for these effects and modeled it by quark transverse momenta in the subprocess. This also allows to generalize
the handbag approach to other amplitudes than the longitudinal ones. Possible infrared singularities are regularized by the quark transverse momenta. 

\begin{figure}[!ht]
\begin{center}
\resizebox{0.37\textwidth}{!}{%
 \includegraphics{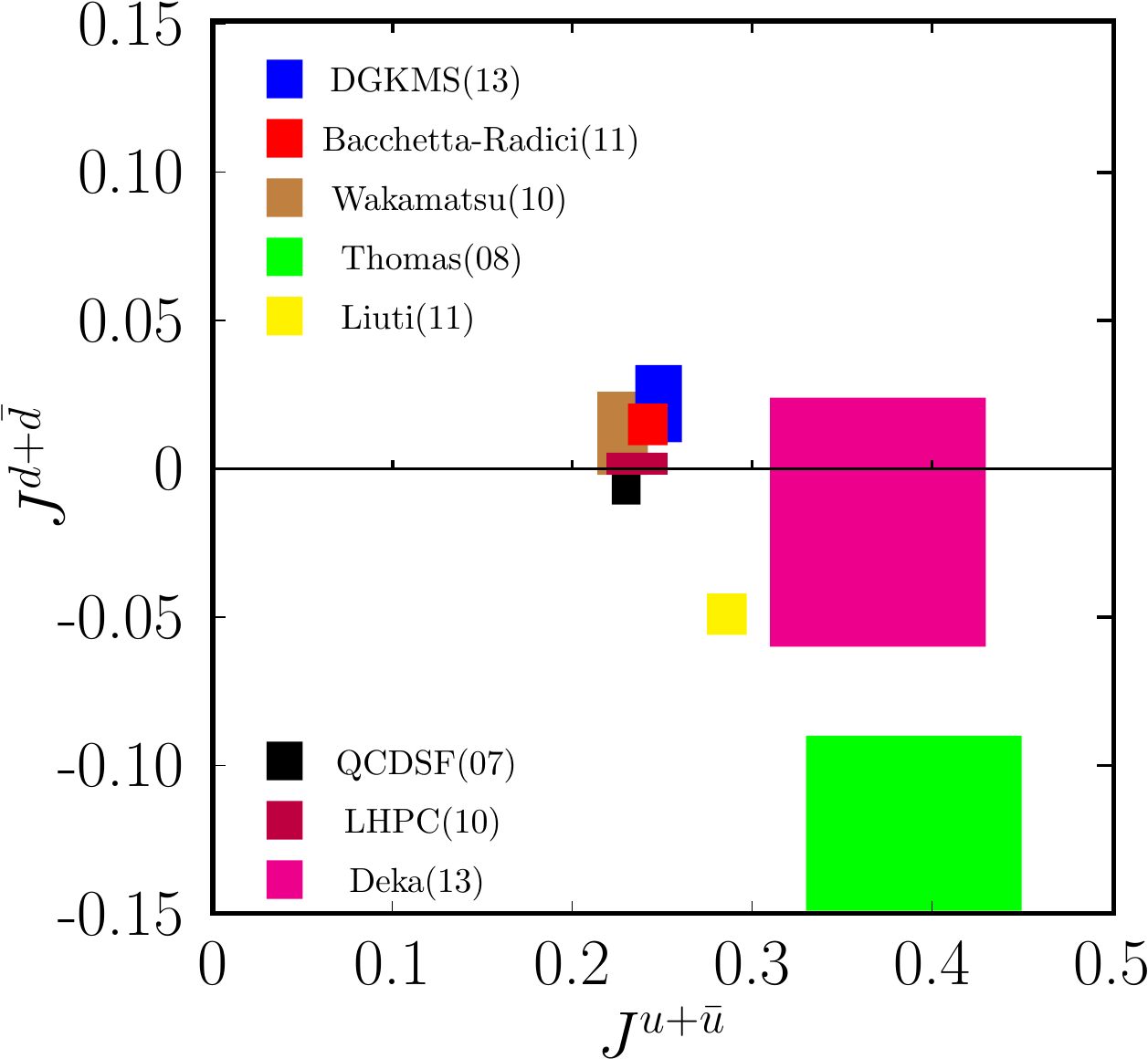}
  }
\caption{Results on the angular momenta for $u$ and $d$ quarks~\cite{KrollarXiv:1410.4450}.} \label{JuJd}
\end{center}
\end{figure}

From an analysis of the nucleon form factors the GPDs $H, E, \widetilde{H}$ for valence quarks have been extracted~\cite{Diehl-KrollarXiv:1302.4604}. The analysis of the longitudinal cross sections for $\rho^0$ and $\phi$ production~\cite{Goloskokov-Krollhep-ph/0611290} provided $H$ for gluons and sea quarks for given valence quark GPD $H$. From the analysis of pion electro-production~\cite{Goloskokov-Kroll1106.4897} one learns about the transversity GPDs $H_T$ and $\bar{E}_T$ for valence quarks. The transversity GPDs go along with a twist-3 pion wave function. This set of GPDs describes as well the spin density matrix elements (SDME) and the spin asymmetry $A_{UT}$ for $\rho^0$ production off a 
transversely. It has also been used to predict other hard exclusive reactions free of parameters (except for possible wave function effects) as for instance the pion-induced exclusive DY process, the SDME and $A_{UT}$ for $\omega$ production or, last not least, DVCS. For the latter two processes good agreement with all small skewness data is found. In combination with a sum rule for the second moments of $E$ and a positivity bound, $A_{UT}$ for DVCS allows for a first estimate of $E$ for gluons and sea quarks (assuming a flavor symmetric sea). With the help of this information the parton angular momenta can be evaluated from Ji's sum rule. The results for $u$ and $d$ quarks are shown in Fig.~\ref{JuJd}.

%
%
\subsection{Modelling GPDs in moment space approach}
\vspace*{-4pt}
\subsubsection*{\hspace*{18pt} \small\em{K.~Kumeri\v{c}ki \& D.~M\"{u}ller}}
\begin{figure*}[ht]
\centerline{\includegraphics[width=0.70\textwidth]{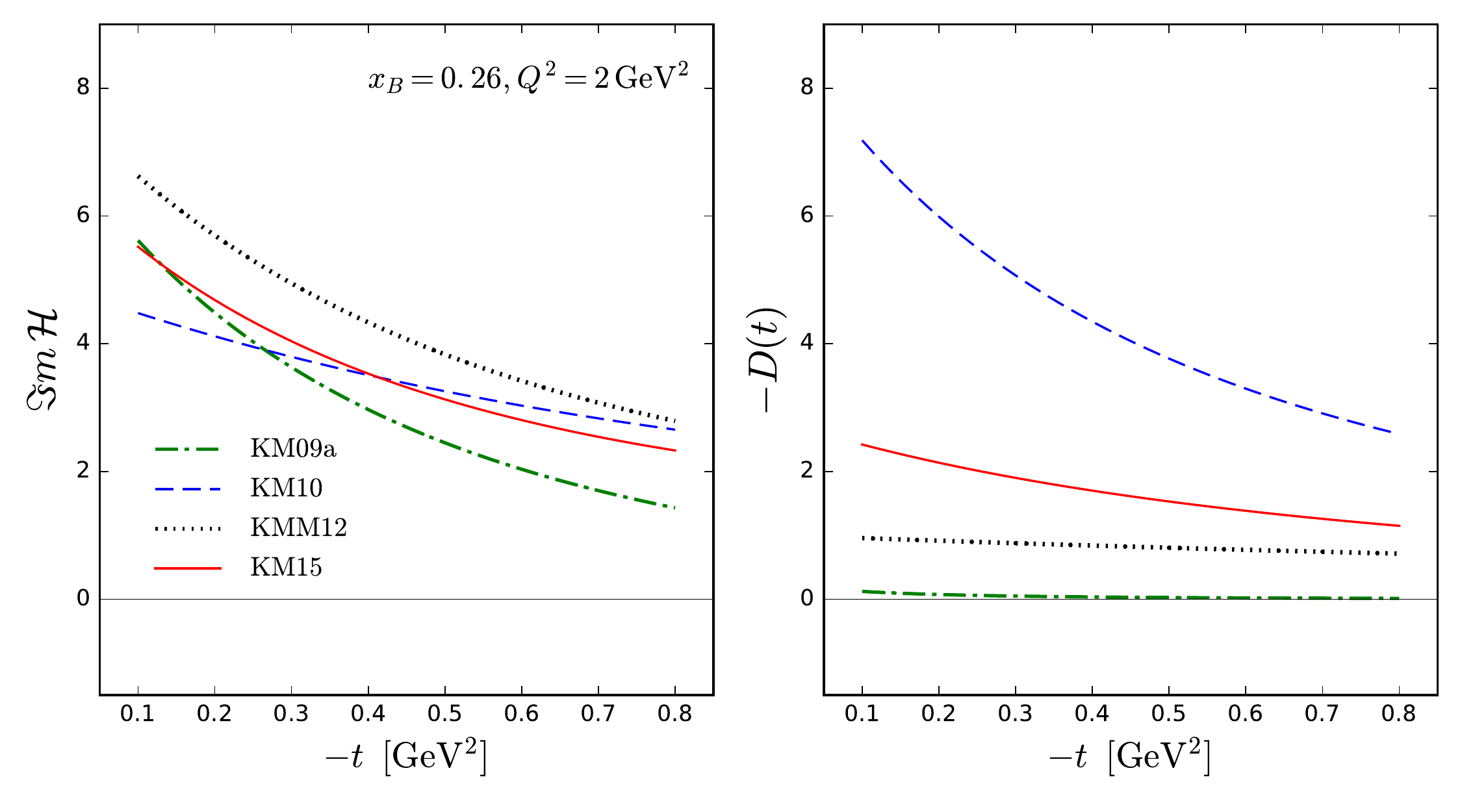}}
\caption{Comparison of size and $t$-dependence of the fitted $\Im{}m\,\mathcal{H}$ CFF (left) and the corresponding $D$-term (right) for several global GPD/CFF fits~\protect\cite{Kumericki:2009uq,Kumericki:2013br,Kumericki:2015lhb}.}
\label{dterm}
\end{figure*}

\vspace*{-4pt}
Global fitting to world DVCS data has been pursued in~\cite{Kumericki:2009uq,Kumericki:2013br,Kumericki:2015lhb}
using a prescription where valence GPDs are modelled mostly on the cross-over $x=\eta$ line, resulting in the imaginary part of Compton Form Factor (CFF). The real part is then obtained using dispersion relations, with separately pa\-ra\-me\-te\-ri\-zed subtraction constant taking the role of the $D$-term. Several fits which all reasonably describe data are plotted on Fig.~\ref{dterm}, where it is visible that DVCS data constrain mostly the $\Im{}m\,\mathcal{H}$, while both the $t$-dependence and absolute size of the $D$-term are quite poorly determined. Availability of DDVCS measurements, providing direct access to the central GPD region $|x|<\eta$, could discriminate among these fits and thus dramatically improve our knowledge of the shape of relevant GPD.

%
%
\subsection{Accessing the real part of the forward $J/\psi$-$p$ scat\-te\-ring amplitude from $J/\psi$ photo-production on protons around threshold}
\vspace*{-4pt}
\subsubsection*{\hspace*{18pt} \small\em{O.~Gryniuk \& M.~Vanderhaeghen}}
\begin{figure*}[h]
\begin{center}
\includegraphics[width=0.36\textwidth]{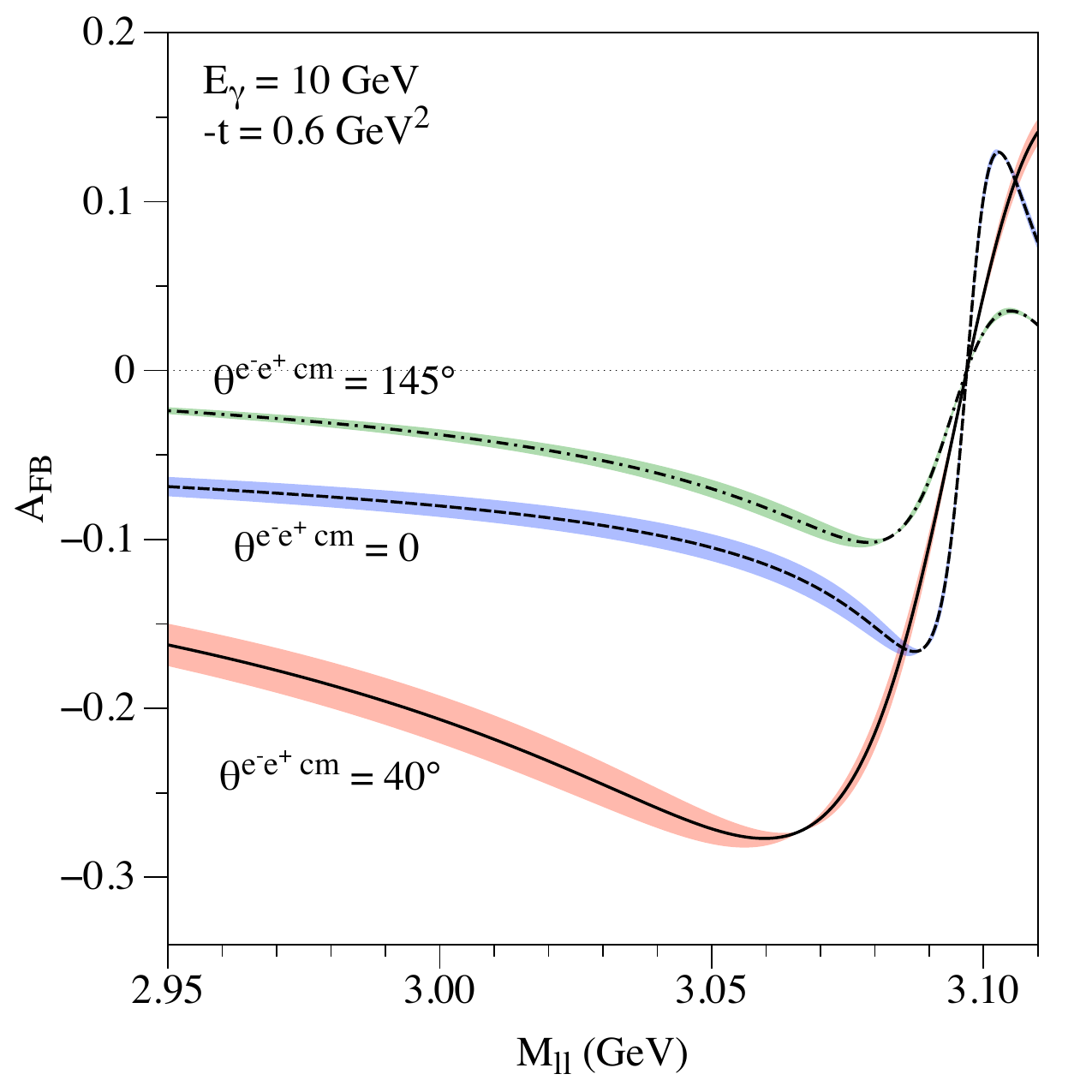}
\includegraphics[width=0.36\textwidth]{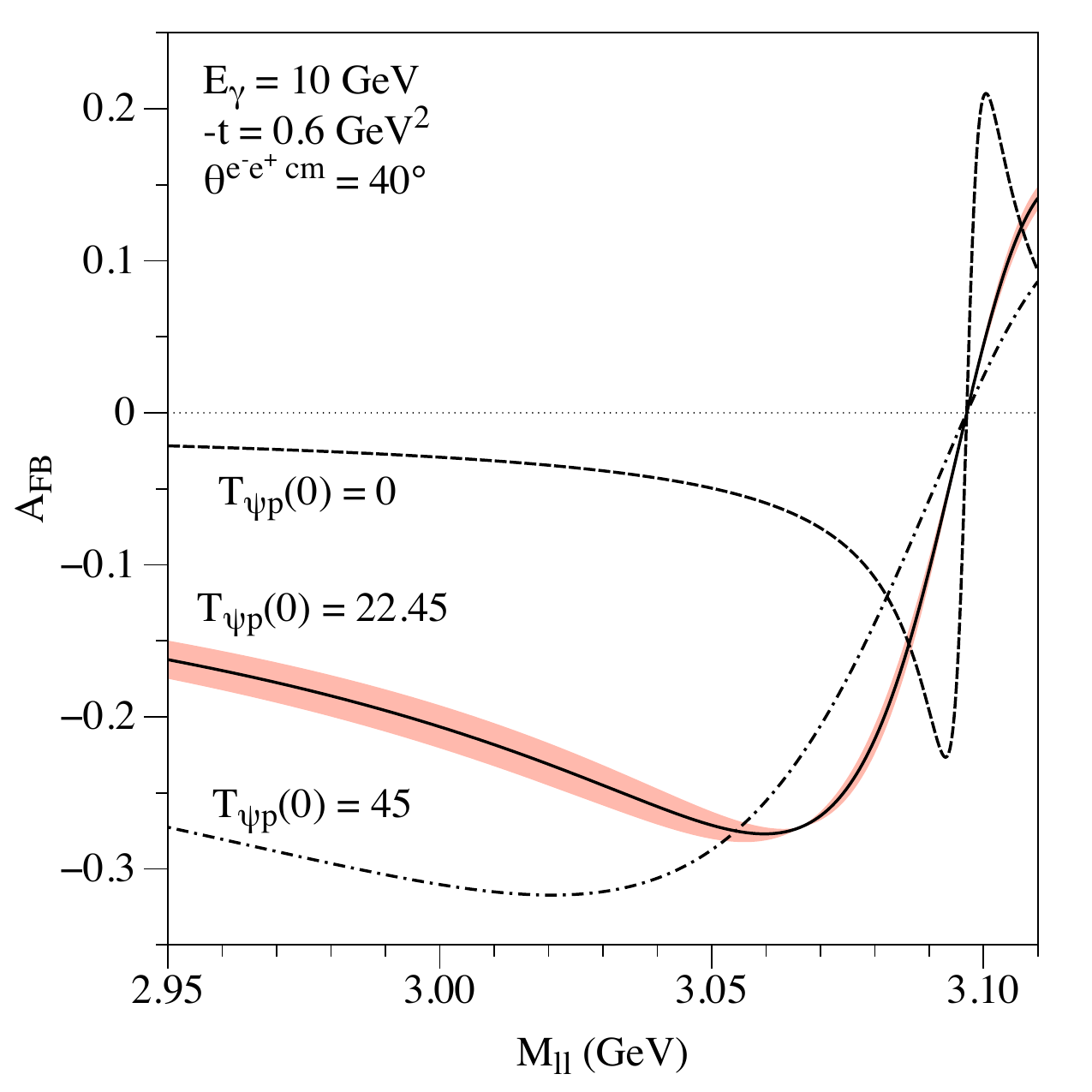}
\caption{Forward-backward asymmetry for the $\gamma \, p \to e^- e^+ \, p$ process as function of the dilepton mass $M_{ll}$ around the $\psi$ resonance: subtraction constant value $T_{\psi p}(0) = 22.45$ (left panel), and $\theta^{e^- e^+}_{cm} = 40^\circ$ (right panel)~\cite{Gryniuk:2016mpk}.}
\label{fig:tcs_ega10_fb}
\end{center}
\end{figure*}

\vspace*{-4pt}
We provide an updated analysis of the forward $J/\psi$-$p$ scattering amplitude, relating its imaginary part to 
$\gamma p \to J/\psi p$ and $\gamma p \to c \bar c X$ cross section data, and calculating its real part through a once-subtracted dispersion relation, see Ref.~\cite{Gryniuk:2016mpk} and references therein. From a global fit to both differential and total cross section data, we extract a value for the spin-averaged $J/\psi$-$p$ $s$-wave scattering length $a_{\psi p} = 0.046 \pm 0.005$~fm, which can be translated into a $J/\psi$ binding energy in nuclear matter of $B_\psi = 2.7 \pm 0.3$~MeV. We estimate the forward-backward asymmetry to the $\gamma p \to e^- e^+ p$ process around the $J/\psi$ resonance, which results from interchanging the leptons in the interference between the $J/\psi$ production and the Bethe-Heitler mechanisms. Fig.~\ref{fig:tcs_ega10_fb} shows that to good approximation this asymmetry depends linearly on $a_{\psi p}$, and can reach values about -25\% for forthcoming $J/\psi$ threshold production experiments at Jefferson Lab. Its measurement can thus provide a very sensitive observable for a refined extraction of $a_{\psi p}$. 

%
%
\subsection{Fits of Compton form factors with time-like Com\-pton scattering off the proton}
\vspace*{-4pt}
\subsubsection*{\hspace*{18pt} \small\em{M.~Boer}}
\begin{figure} 
\begin{center}
\includegraphics[width=0.6\textwidth, page=1, angle=270]{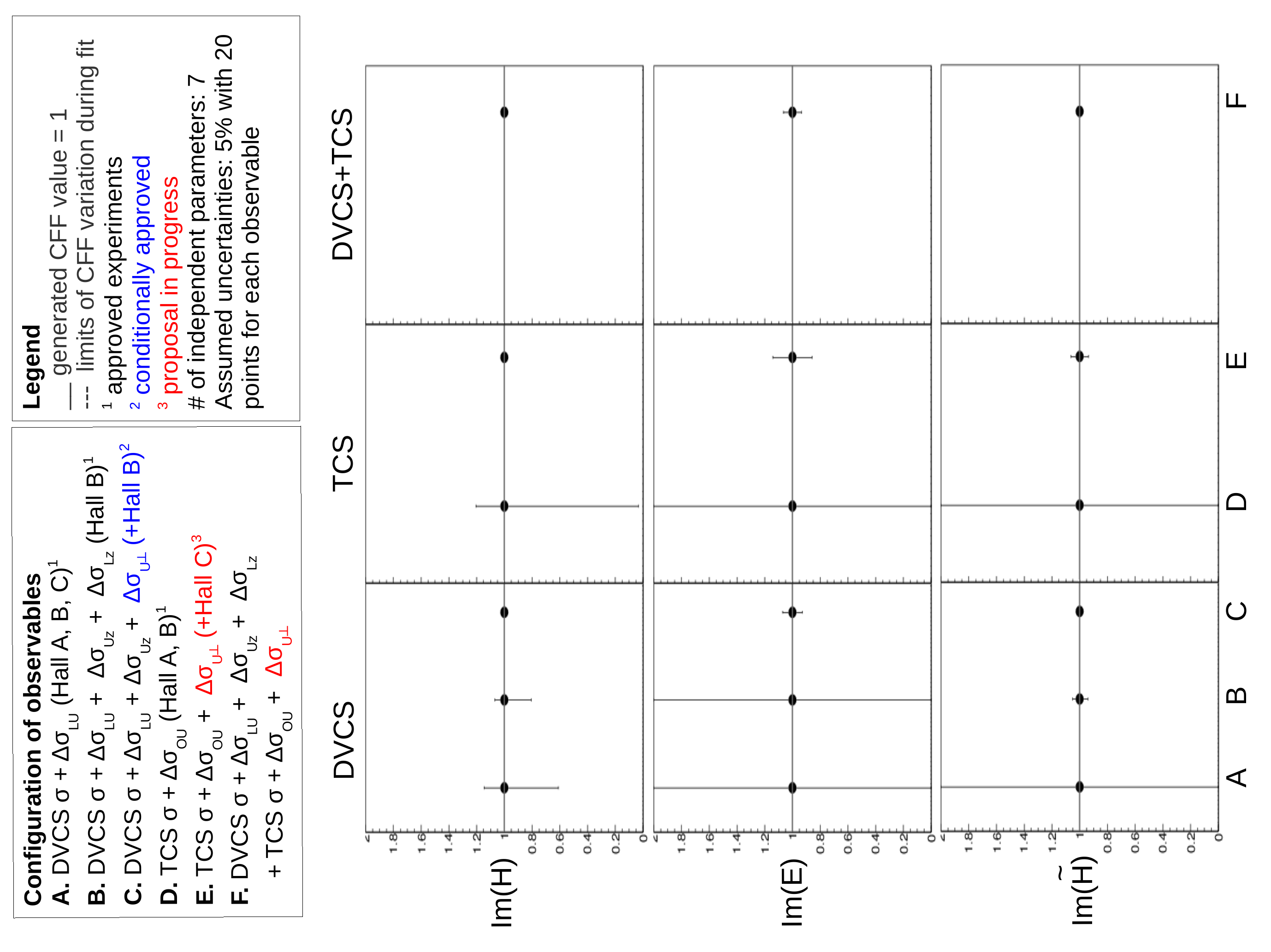}  
\includegraphics[width=0.6\textwidth, page=2, angle=270]{Fits_newfig_zoom.pdf}  
\caption{Fit results for imaginary part of CFFs from various sets of DVCS (left), TCS (center) and DVCS+TCS (right) observables. Simulations assume 5\% uncertainties for all observables which correspond to approved or proposed measurements  at Jefferson Lab.}
\label{fig:fits}
\end{center} 
\end{figure}

\vspace*{-4pt}
We performed fits of CFFs from TCS simulated observables~\cite{Boe15-1}, in a kinematic domain that is accessible by future experiments at Jefferson Lab. Selecting a kinematic bin accessible from both DVCS and TCS experiments, it is possible to evaluate the sensitivity of each single reaction channel to the CFFs. The combination of observables measured with both processes further demonstrate the benefit of measuring TCS. \\ 
The approved experiments at Jefferson Lab (CLAS12 and SoLID) will mostly allow for  the extraction of the imaginary part of the CFFs $\mathcal{H}$, $\mathcal{E}$, and $\widetilde{\mathcal{H}}$. Compared to DVCS, the sensitivity of TCS to GPDs is weaker. However, the comparison of results from these two processe is essential~\cite{Die03} for a first experimental proof of GPDs universality. Combined fits of DVCS+TCS demonstrate an improvement of the GPD knowledge corresponding to the addition of independent constrains from TCS quasi-data. Fig.~\ref{fig:fits} summarizes the comparison of the CFFs sensitivity from future JLab experiments with some DVCS, TCS and DVCS+TCS.
 
%
%
\subsection{Partons: \bf{PAR}tonic \bf{T}omography \bf{O}n \bf{N}ucleon \bf{S}oftware}
\vspace*{-4pt}
\subsubsection*{\hspace*{18pt} \small\em{L.~Colaneri}}
\vspace*{-4pt}
GPDs are 3-D functions describing the partonic structure of nucleons. Directly related to matrix elements of the QCD energy-momentum tensor, they hold information on the longitudinal momentum distribution and transverse position of partons. A consistent set of data is already available from DVCS and DVMP and a lot more is awaiting from future experiments at JLab, COMPASS, EIC and possibly more experimental sites. \\ 
PARTONS~\cite{Ber15} (PARtonic Tomography On Nucleon Software) is a C++ software framework dedicated to the phenomenology of Generalized Parton Distributions. PARTONS provides a necessary bridge between GPD models and experimental data measured in various exclusive channels. This framework, currently under construction, will be useful not only for theorists to develop  new models but also to interpret existing measurements and even design new experiments.

%
%
\section{Experimental projects}

The detection of lepton pairs is a new powerful tool complementing the current GPD program mostly focusing on the DVCS. In this workshop we explored the double deeply virtual Compton scattering, the time-like Compton scattering, the Drell-Yan reaction, and exclusive meson production.

%
%
\subsection{DDVCS}

\subsubsection{The CLAS12 project}
\vspace*{-4pt}
\subsubsection*{\hspace*{18pt} \small\em{N.~Baltzell}}
\vspace*{-4pt}
A DDVCS experiment using a modified CLAS12 detector and 11 GeV electron beam in Hall B was intoduced as a letter of intent to Jefferson Lab's PAC44~\cite{Ste16}. The di-muon final state is employed to avoid anti-symmetrization issues for DDVCS, as well as reducing combinatorial background in a simultaneous measurement of $J/\Psi$ electro-production.\\
The proposed experiment uses standard CLAS12 forward detectors, but shielded by a 30~cm thick tungsten absorber to mitigate electromagnetic and hadronic backgrounds and allow operation as a muon detector, while CLAS12 central detectors are removed.  This also permits a factor of 100 increase in luminosity relative to the baseline CLAS12 setup by reducing the occupancies in the forward drift chambers. Upstream of the forward shielding, the setup is supplemented by a new PbWO$_4$ calorimeter for electron detection and a GEM tracker for improved vertex and momentum reconstruction, and background rejection. Simulations confirm manageable rates and occupancies at a luminosity of $10^{37}$cm$^{-2} \cdot$s$^{-1}$, including few-\% drift chamber occupancies, a di-muon trigger with 360 Hz event rate, and background in final offline reconstruction of DDVCS exclusivity below 20\%.  Expected sensitivity for 100 days of running includes a first measurement of sign flip in DDVCS beam spin asymmetry betwen space-like- and time-like-dominated regimes with $Q^2 \in (2 - 3)$~GeV$^2$, $-t \in (0.1 - 0.4)$~GeV$^2$, and $x_B \in (0.12 - 0.22)$, and varying $Q{^\prime}^2 \sim$~1.2, 2.0, 2.8, 3.6~GeV$^2$. A simultaneous measurement of near threshold $J/\Psi$-electro-production cross sections, decay angular distributions, and $\sigma_L/\sigma_T$ assuming $s$-channel helicity conservation, covers $W \in (4.1 - 4.5)$~GeV and $-t \in (0.5 - 4.5)$~GeV$^2$ for $Q^2$ up to 
2.5~GeV$^2$.
\begin{figure}[ht!]
\centering
    \includegraphics[width=0.45\textwidth, height=5.9cm]{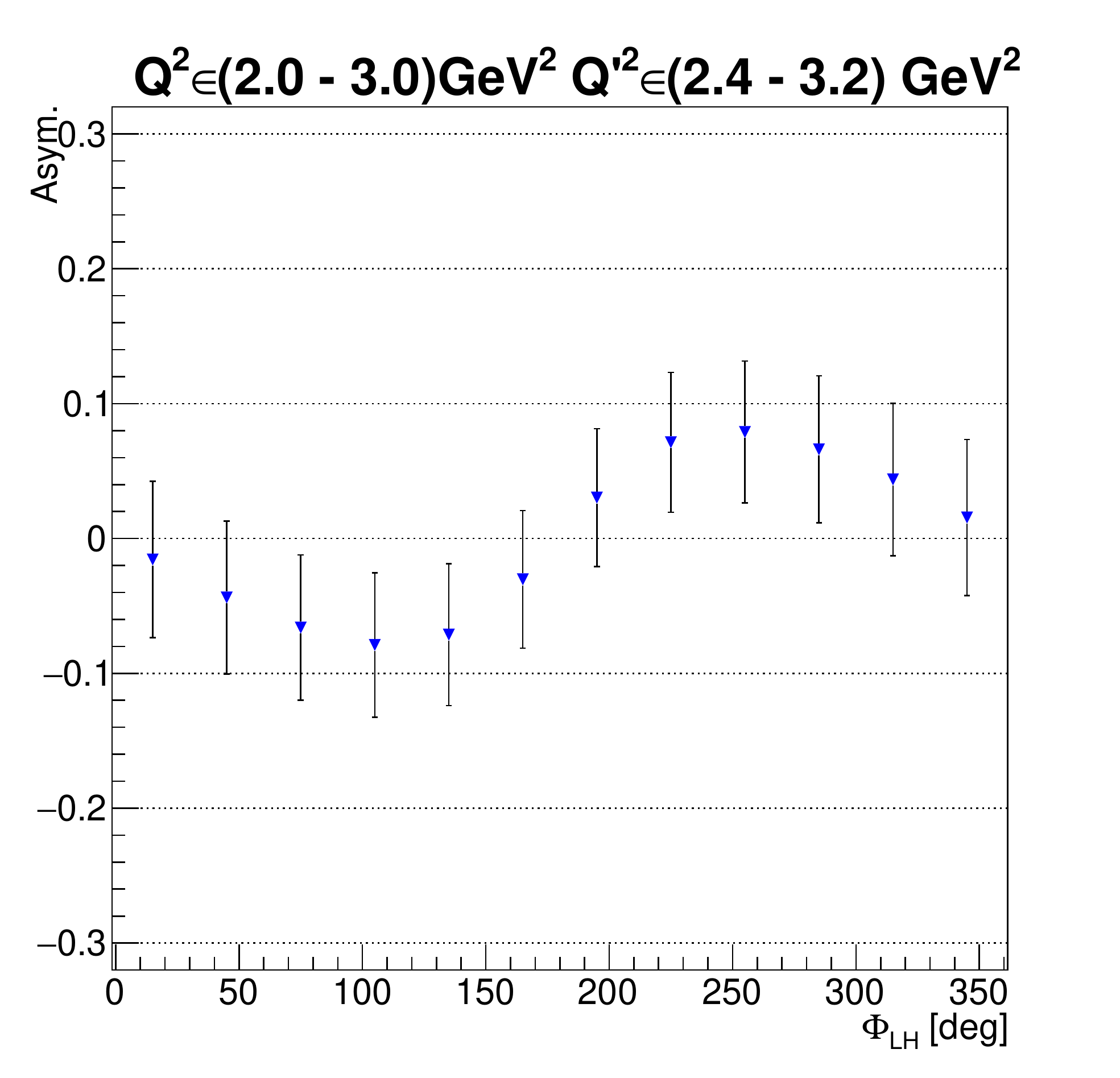}
    \caption{Projected statistical uncertainties, based on Bethe-Heitler cross section, on the beam spin asymetry calculated from the VGG model, for one kinematics example of the proposed CLAS12 DDVCS experiment~\cite{Ste16}.}
\end{figure}

\subsubsection{The SoLID project}
\vspace*{-4pt}
\subsubsection*{\hspace*{18pt} \small\em{A.~Camsonne \& K.~Gnanvo \& E.~Voutier}}
\begin{figure}[t!]
\begin{center}
\includegraphics[width=0.485\textwidth]{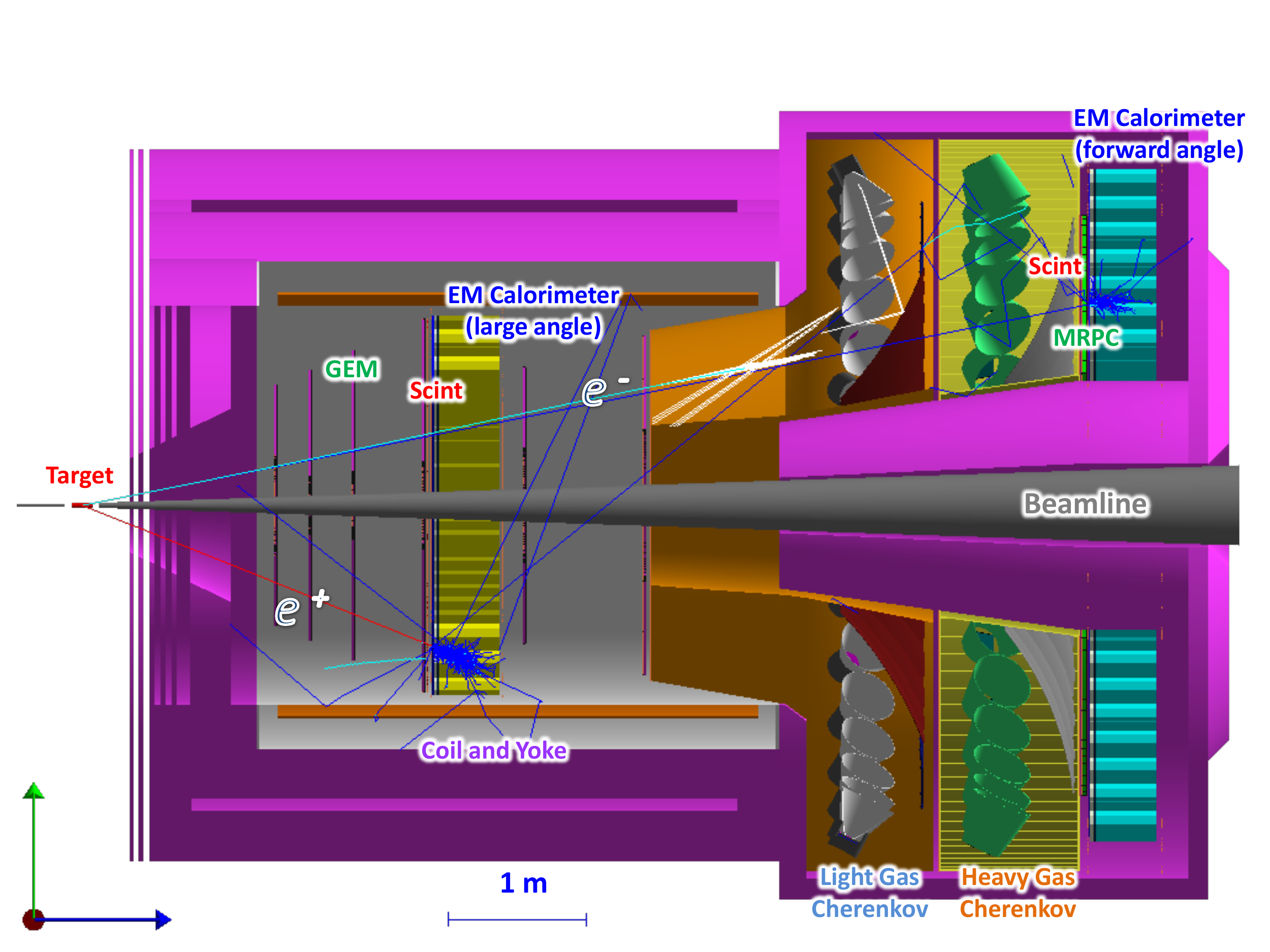}
\caption{SoLID J/$\Psi$ configuration with 15~cm long hydrogen target desinged to achieve a 10$^{37}$cm$^{-2}\cdot$s$^{-1}$ luminosity. \label{fig:SoLIDJPsi}}
\end{center}
\end{figure}

\vspace*{-4pt}
The SoLID spectrometer (Fig.~\ref{fig:SoLIDJPsi}) is a large acceptance detector embedded into a solenoidal magnetic field and designed to run at high luminosity. Among the experimental program attached to this detector, the study of the TCS  reaction and the $J/\Psi$ production are involving di-lepton detection. A DDVCS parasitic measurement during the $J/\Psi$   experiment was proposed at Jefferson Lab's PAC43~\cite{Vou15}. It implies the development of specific large area muon detectors to be placed inside the iron yoke of the SoLID magnet as well as outside and behind the magnet for a full angle coverage. \\
Micro Pattern Gaseous Detector (MPGD) technologies are the ideal candidate to achieve large area position detectors with high rate capability and excellent spatial resolution ($<100$~$\mu$m). Gas Electron Multiplier (GEM) and Micromesh Gaseous  Structure (Micromegas) are well established MPGD technologies used as muon tracking systems in current and future high energy physics and nuclear physics experiments. The development of low cost MPGD technologies with improved performances for  specific applications, particularly the DDVCS-SoLID experiment, is required. The resistive micro-well ($\mu$-RWELL) detector is a new concept combining the recent progresses of both GEM and Micromegas into one MPGD technology. It also offers a very simple, straightforward and low cost approach for the detector assembly compare to triple-GEM or Micromegas chambers. Chromium GEM (Cr-GEM) is another approach based on very low mass detectors by replacing the 5~$\mu$m copper layer used as electrode for GEM foils by 100~nm chromium leading to a reduction by a factor 2 of the total material in a triple-GEM detector. An aggressive muon detector R\&D program is ongoing to meet the requirements for DDVCS with SoLID.

\begin{figure}[!t]
\centering
    \includegraphics[width=0.45\textwidth]{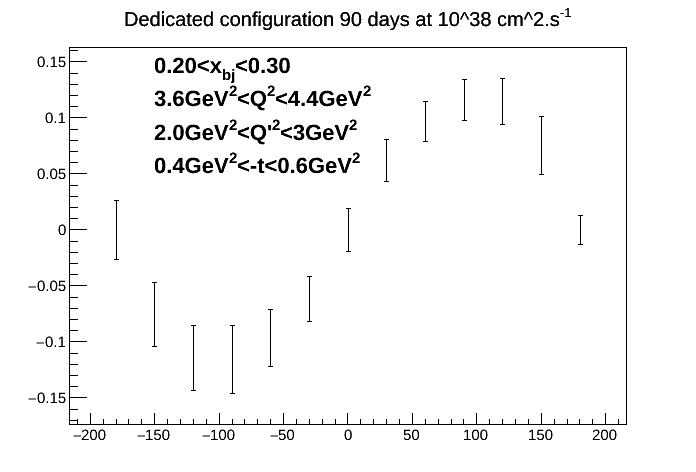}
    \caption{Projected statistical uncertainties from the VGG model, for the proposed SoLID DDVCS dedicated experiment 
    for 90 days data taking at $10^{38}$cm$^{-2} \cdot$s$^{-1}$ luminosity~\cite{Vou15}. \label{fig:DDVCSDed}}
\end{figure}

A dedicated setup would allow a better reach in $Q{^\prime}^2$ by moving the target inside the solenoid to improve angular  acceptance at large angle. The dedicated setup could involve additionnal absorber in front the SoLID calorimeter degrading the energy resolution but improving the rate capability and radiation hardness, the resolution on the final momentum being  determined from the GEM trackers. Such a setup is being studied to run at $10^{38}$cm$^{-2} \cdot$s$^{-1}$ luminosity and  would allow to scan a higher range in $x_{B}$ and $Q^2$ (Fig.~\ref{fig:DDVCSDed}).

%
%
\subsection{TCS}

\subsubsection{Prospects for observables with proton and neutron targets}
\vspace*{-4pt}
\subsubsection*{\hspace*{18pt} \small\em{M.~Boer}}
\vspace*{-4pt}

\begin{figure}[!b]  
  \begin{center}
     \includegraphics[width=0.425\textwidth]{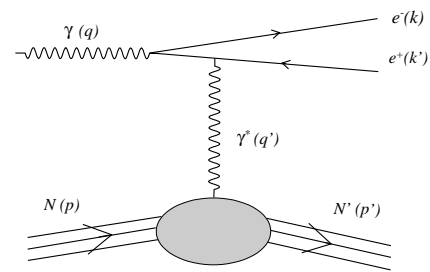} 
\vspace*{-4pt}
    \caption{The Bethe-Heitler processfor in the dilepton photo-production channel (the crossed diagram is not represented).}
    \label{fig:TCSBH}
  \end{center} 
\end{figure}
\begin{table*}[!th]
{\footnotesize
\begin{tabularx}{\textwidth}{|X|X|X|X|}
\hline
 \bf{ Observable (proton target)} & \bf{ Experimental challenges} &  \bf{Main GPDs interest}  &  \bf{Future experiments at JLab 12 GeV } \\ \hline 
    Unpolarized cross section & 1 or 2 order of magnitude lower than DVCS, requires high luminosity  & $\Re e(\mathcal{H}$), $ \Im m (\mathcal{H}$)  & Approved proposals for Hall A and B \\ \hline
Circularly polarized beam spin asymmetry & Easiest observable to measure. & $\Im m(\mathcal{H}$), $\Im m(\widetilde{\mathcal{H}}$). Sensitivity to quark angular momenta, particularly for neutron & Approved proposals for Hall A and B
 \\ \hline
    Linearly polarized beam spin asymmetry & Need high luminosity, at least 10 times larger than for the circularly polarized beam. Need electron tagging if using a quasi-real photon beam. & $\Re e (\mathcal{H}$), $D$-term. Good to discriminate models and very important to bring constrains on the real part of CFFs. & None \\ \hline
    Longitudinaly polarized target single spin asymmetry & Polarized target & $\Im m(\widetilde{\mathcal{H}}$) & None
    \\ \hline
    Transversely polarized target single spin asymmetry & Polarized target and high luminosity required & $\Im m (\widetilde{\mathcal{H}}$), $\Im m(\mathcal{E}$) & LOI in Hall C
    \\ \hline 
    Double spin asymmetries with a circularly polarized beam and a polarized target & Polarized targets, very high luminosity 
    and precision are requested.
    & Real part of all CFFs & None
     \\ \hline 
    Double spin asymmetries with a linearly polarized beam and a polarized target & Polarized target, electron tagging, very high luminosity, and precision are requested.
     & Les interesting observables, bring almost the same information than single target spin asymmetries & None 
     \\ \hline
     All these observables for a 
     neutron target & Need 1 or 2 order of magnitude higher luminosity Larger target spin asymmetries and smaller beam spin asymmetries are expected.
     & Flavor decomposition, sensitivity to quark angular momenta & None
     \\   \hline
 \end{tabularx}
}
\caption{Summary of TCS observables, their interest for GPD physics, and the attached experimental challenges. The third  
column indicates the CFFs sensitivity~\cite{Boe16}.}
\label{table:tcsgpd}
\end{table*}

The TCS process (Fig.~\ref{fig1-diag}) interfers with the Bethe-Heitler process (Fig.~\ref{fig:TCSBH}) which dominates the cross section in the JLab kinematic region. The cross sections for the different combinations of beam and/or target spin asymmetries are discussed~\cite{Boe15}. We emphasize the fact that as for DVCS measurements, TCS is a direct process to access GPDs and brings invaluable information for GPD extraction. For instance, while single spin asymmetries (target or circularly polarized photons) are sensitive to the imaginary part of amplitudes, double spin asymmetries with circularly polarized photons and  single spin asymmetries with linearly polarized photons are sensitive to their real part. The latter observables are more difficult to access experimentaly, particularly because of luminosity constraints, but are very important for GPDs understanding especially regarding the CFFs real part. An additional difficulty in measuring these observables is the fact that the BH contribution does not cancel, contrary to asymmetries sensitive to the imaginary part of the amplitudes. The Tab.~\ref{table:tcsgpd} summarizes the various observables, their interest in constraining CFFs, the foreseen experimental difficulties, and the current experimental status at Jefferson Lab. 

%
%

\subsubsection{Time-like Compton scattering with CLAS12}
\vspace*{-4pt}
\subsubsection*{\hspace*{18pt} \small\em{S.~Stepanyan}}
\begin{figure}[!t]
\begin{center}
\includegraphics[width=0.425\textwidth]{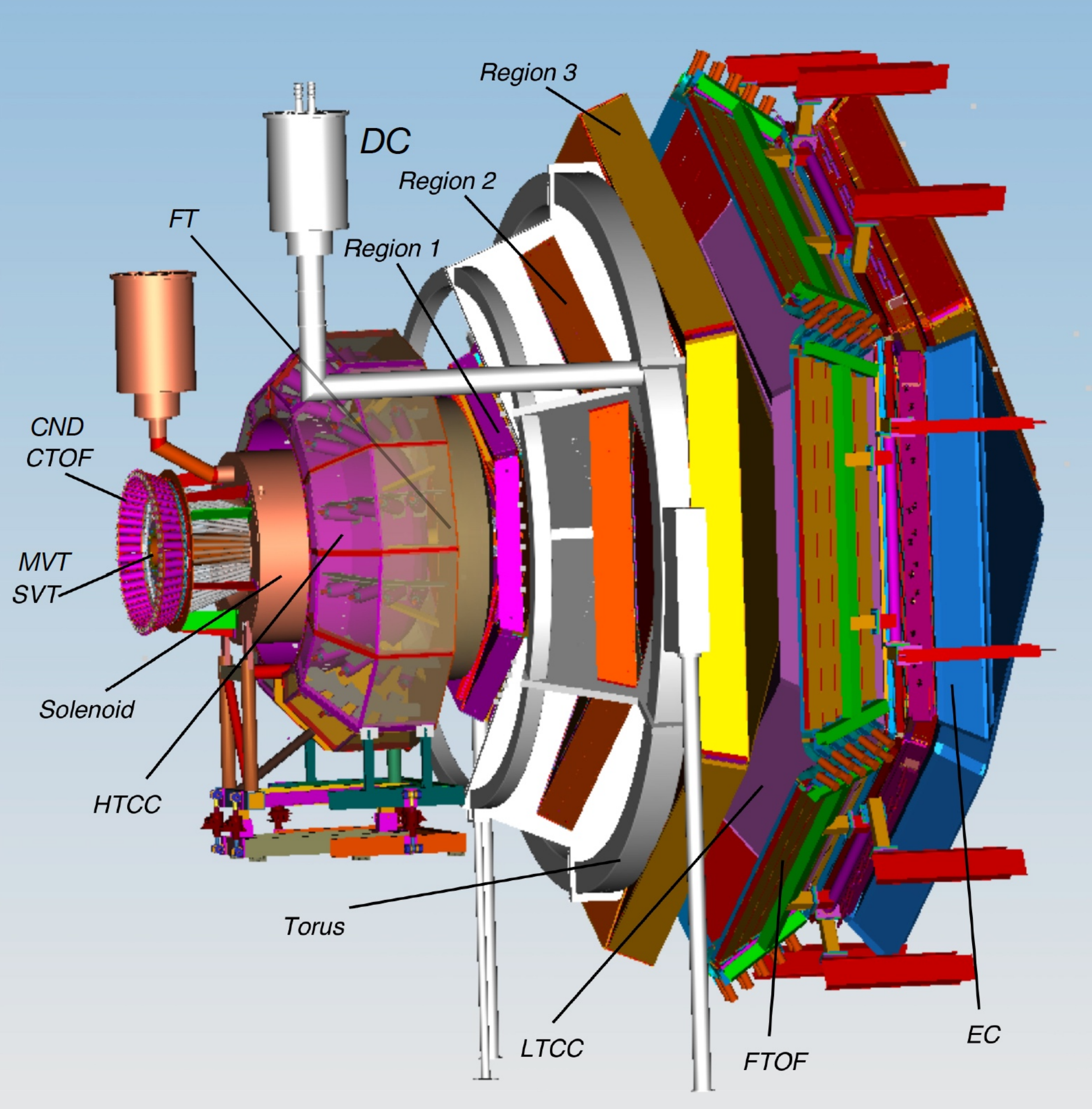}
\caption{The CLAS12 detector in experimental Hall B at Jefferson Lab.}
\label{fig:clas12}
\end{center}
\end{figure}

\vspace*{-4pt}
A new detector, CLAS12, is being commissioned in experimental Hall~B with up to 11~GeV electron beams from the upgraded CEBAF machine at Jefferson Lab. CLAS12 is a large acceptance, multi-purpose detector capable of detecting and identifying neutral and charged particles in the full range of available momentum space. One of the key characteristics of the detector is its high luminosity, $\mathcal{L}$=$10^{35}$~cm$^{-2}\cdot$s$^{-1}$, an important parameter for executing a physics program using the exclusive reactions.  The CLAS12 detector (Fig.\ref{fig:clas12}) has two parts, the forward detector (FD) and the central detector (CD). The FD is based on a six-coil superconducting toroidal magnet and includes micromegas vertex tracker (MVT), three regions of drift chambers (DC), high and low threshold \v{C}erenkov counters (HTCC and LTCC), scintillation counters (FTOF), and the electro-magnetic calorimeters (EC). In the very forward region (from $2^\circ$ to $4.5^\circ$) FD is complemented with a forward tagger system that will detect electrons and photons in the full momentum range. The CD is based on a $5$ T superconducting solenoid magnet. It includes Silicon and Micromegas Trackers (SVT, MVT), scintillation counters (CTOF), and the central neutron detector (CND). The approved physics program on CLAS12 covers a wide range of studies of meson and baryon spectroscopy, nucleon and nuclear structure, quark propagation, and hadronization.

\begin{figure}[!t]
\begin{center}
\includegraphics[width=0.45\textwidth]{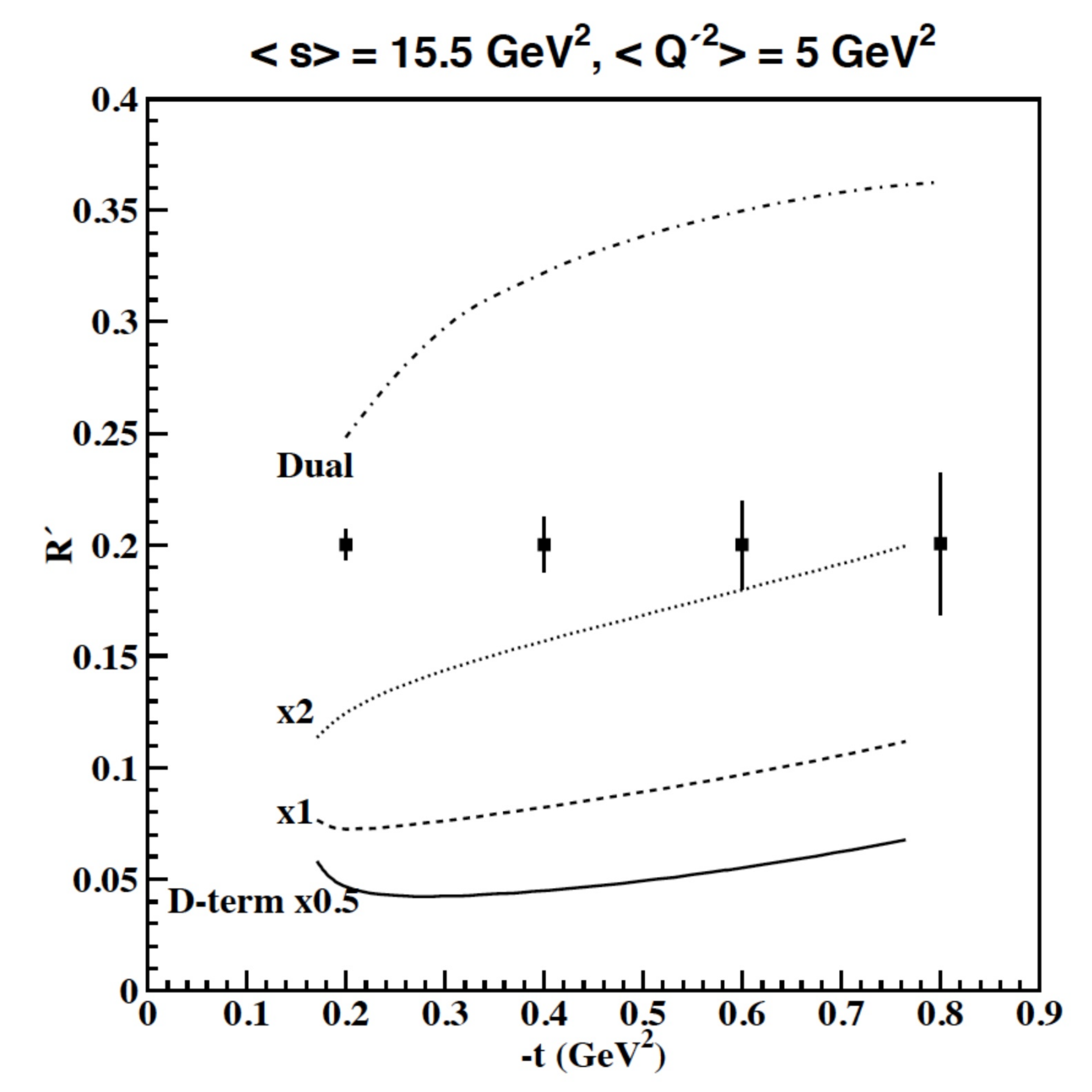}
\caption{Expected results from CLAS12 ($e^+e^-$) quasi-real photo-production measurements on the $\cos(\phi_{c.m.})$ moment of lepton decay angular distribution in one time-like photon kinematic bin as a function of transferred momentum squared, shown together with predictions of the dual parametrization model~\cite{{dual-1},{dual-2}} and double-distribution model~\cite{double} with different weights applied to the $D$-term contribution.}
\label{fig:tcs1bin}
\end{center}
\end{figure}

The time-like Compton scattering is one of the key reactions for the CLAS12 GPD program and will run together with other CLAS12 experiments that use a liquid hydrogen target and an $\sim 11$ GeV electron beam. The photo-production of the electron-positron pairs will be studied in the reaction $ep~\to~p^\prime e^+ e^- (e^\prime)$ where the recoil proton and decay leptons of the time-like photon will be detected in CLAS12, while the kinematics of the scattered electron will be deduced from the missing momentum analysis. In such a setting, due to the forward peaking nature of the electron scattering cross section, the ($p^\prime e^+ e^-$) are produced in  quasi-real photo-production kinematics, Q$^2=-(q^\prime-q)^2\sim0$ GeV$^2$ (here $q$ and $q^\prime$ are four-momenta of the incoming and scattered electrons, respectively). The proposed measurements will be carried out in the resonance-free region of the lepton pair invariant mass ($M^2_{e^+e^-}\equiv Q^{\prime2}$) between the $\rho^\prime$ and the J/$\psi$. The differential cross sections, and the azimuthal angular distributions of the lepton pair will be studied in a wide range of four-momentum transfer $-t$ and center-of-mass (c.m.) energy $s$. With unpolarized photon beams, the cosine moments of the weighted and $\theta_{c.m.}$-integrated cross section will allow to measure the real part of the Compton amplitude, not accessible directly in electron-DVCS measurements, and gives access to the $D$-term in the GPD parametrization. The measurements with circularly polarized photons, available for all CLAS12 electron runs (electron beam is always longitudinally polarized), gives access to the imaginary part of CFFs and allows direct comparison with DVCS measurements to test GPDs universality. In Fig.~\ref{fig:tcs1bin}, the $\cos(\phi_{c.m.})$ moment of the decay lepton angular distribution in one time-like photon kinematic bin as a function of transferred momentum squared is shown together with predictions of the dual parametrization~\cite{{dual-1},{dual-2}} and the double-distribution~\cite{double} models. Different weights have been applied to the $D$-term contribution in the later calculations. The statistical errors correspond to 100 days of running with the CLAS12 design luminosity. Similar sensitivity will be reached in multiple bins of $s$- and $Q^{\prime 2}$-variables.    


%
%
\subsubsection{Time-like Compton scattering at SoLID}
\vspace*{-4pt}
\subsubsection*{\hspace*{18pt} \small\em{Z.W.~Zhao}}
\vspace*{-4pt}
SoLID TCS using the $J/\psi$ experimental setup Fig.~\ref{fig:SoLIDJPsi} is the perfect next stage experiment after the CLAS12 TCS. 
It will provide the essential cross-check by using two very different detectors to measure the same process, a mandatory approach since TCS is still a new tool for GPD studies. The experiment will measure TCS cross sections and beam spin asymmetries in a wide range of outgoing photon virtuality ($4<Q'^2<9$~GeV$^2$) and skewness ($0.1<\xi<0.35$). Taking advantages of SoLID's high luminosity capability, it will collect unprecedented amount of high quality data and push TCS study to the precision era. Besides testing GPDs universality by comparing to DVCS, this experiment has the potential to observe next-to-leading order effects of GPDs. These measurements will provide critical input to the GPD global fitting.

%
%
\subsubsection{Time-like Compton scattering project in Hall C at JLab}
\vspace*{-4pt}
\subsubsection*{\hspace*{18pt} \small\em{V.~Tadevosyan}}
\vspace*{-4pt}
A project for the measurement of TCS off transversely polarized proton in Hall C at JLab is developping~\cite{Tad15}. The projected experimental setup allows for the coincident detection of recoil protons by the time-of-flight technique and the $e^+e^-$ pair decay by energy measurements. The quasi-real scattering of 11 GeV incident electron is identified by means of reconstruction of missing 4-momentum. The VGG GPD model based TCS calculations~\cite{{Boe15-1},{Boe15-2}} predict target (TA), beam (BA) and double (DA) spin asymmetries significant for reliable detection, and sensitivity to imaginary parts of $\mathcal{H}$ and $\mathcal{E}$ CFFs (TA), and to the real part of the TCS amplitude (DA). Phase space coverage, distributions of the TCS kinematic variables and count rates are obtained from acceptance simulations of the setup. The proposed measurement is complementary to JLab approved E12-12-001~\cite{Nad12} and E12-12-006A~\cite{Zha15} experiments in Halls B and A respectively.
The proposed experimental setup consists of a transversally polarized target, pairs of fiber trackers, scintillator hodoscopes, and PbWO$_4$ calorimeters. The phase space coverage of the proposed measurements onto the ($Q^{'2}, \xi$) plane will allow for studies of the $Q^2$-, $\xi$- and $t$-de\-pen\-den\-ces. 

%
%
\subsection{Drell Yan}

\subsubsection{Nucleon structure via recent Drell-Yan ex\-pe\-ri\-ments}
\vspace*{-4pt}
\subsubsection*{\hspace*{18pt} \small\em{M.~Boer}}
\vspace*{-20pt}
\begin{figure}[h!]
\begin{center}
\includegraphics[width=0.485\textwidth]{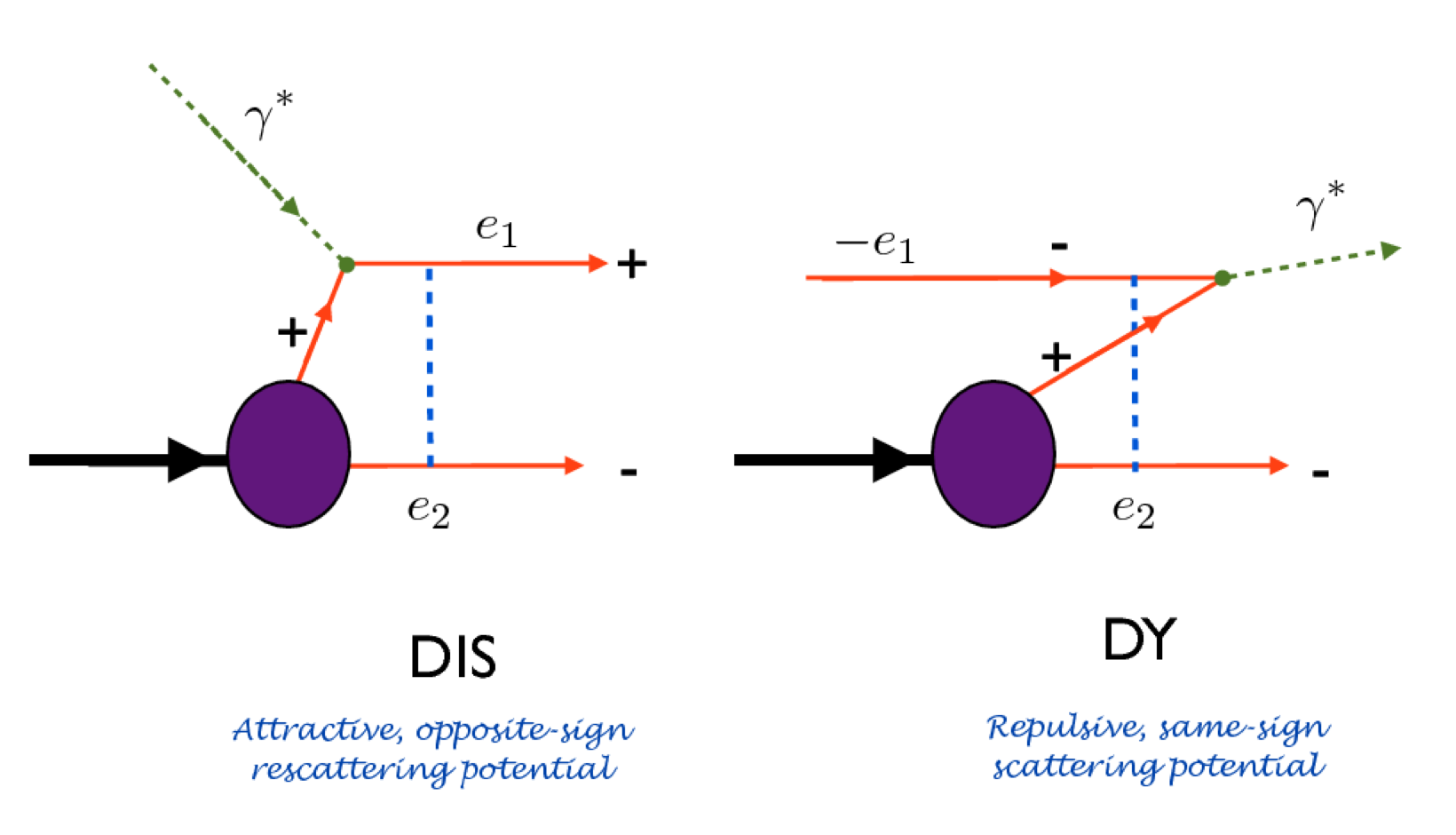}  
\caption{Illustration of Sivers effect from initial state versus final state interaction in DY and DIS 
reactions~\cite{Brodsky:2013oya}.}
\label{fig:DY}
\end{center} 
\end{figure} 
The measurement of the DY reaction is of particular interest to access TMDs, especially the Sivers TMD~\cite{{Sivers:1989cc},{Collins:2002kn}}. The latter correlates the parton longitudinal momentum with their transverse momentum distribution. Compare to Deep Inelastic Scattering (DIS) experiments, it is expected to measure spin asymmetries from Sivers effect 
with an opposite sign in DY (Fig.~\ref{fig:DY}. The Sivers effect and this sign change is understood as coming from attractive final  state interaction in DIS as compared to repulsive initial state interaction in DY. Several dedicated experiments have been developed with a transversely polarized target.The COMPASS experiment took data in 2015 with a totaly new dedicated setup~\cite{Aghasyan:2017jop}. The FERMILAB collaboration, which  has a long history in measuring DY proposed recently a dedicated experiment with a transversally polarized target~\cite{Brown:2014sea}. The PHENIX collaboration is also taking advantage of transversally polarized proton-proton 
and proton-ion data accumulated in 2015 to contribute to this physics~\cite{Aschenauer:2013woa}.

%
%
\subsubsection{Pion Drell-Yan at J-PARC}
\vspace*{-4pt}
\subsubsection*{\hspace*{18pt} \small\em{W.C.~Chang}}
\begin{figure}[h!]
\begin{center}
\vspace*{-7pt}
\includegraphics[width=0.425\textwidth]{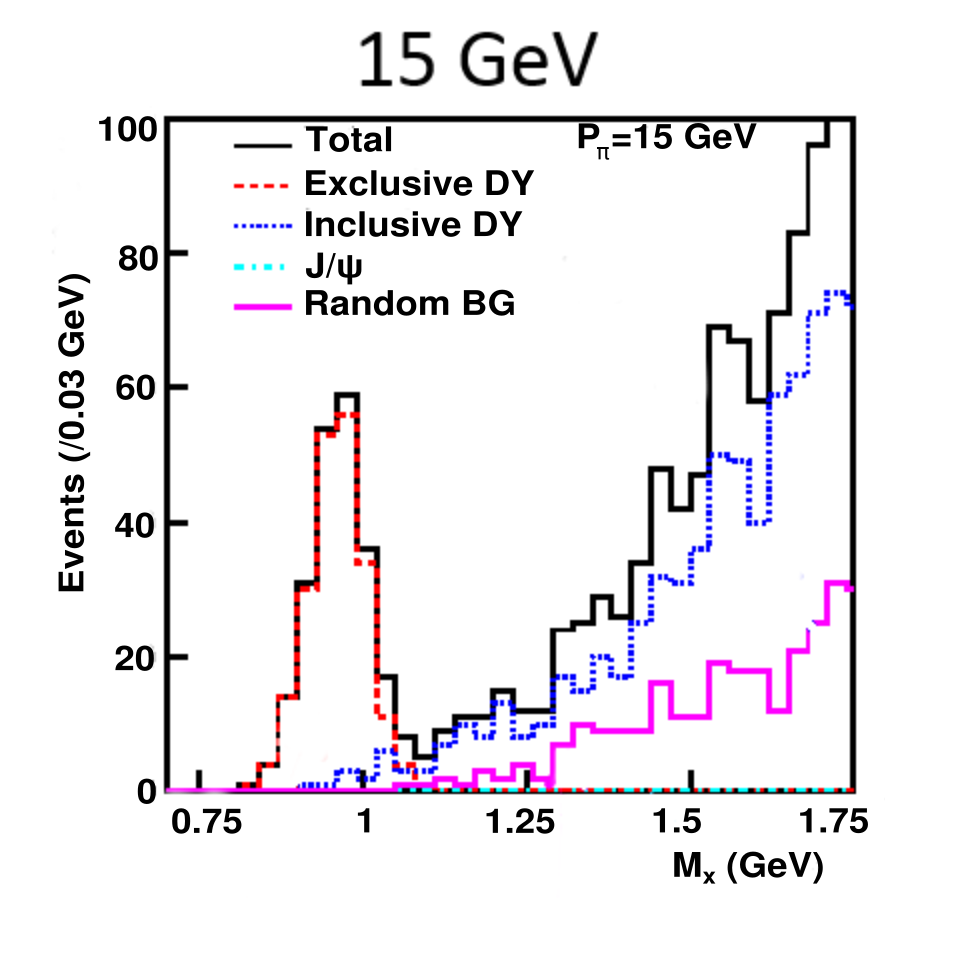}
\vspace*{-7pt}
\caption{Estimation of missing mass plots of DY at 15~GeV incident pion energie at J-PARC for 50 days of beam time. One can clearly see the separation between the exclusive and inclusive DY processes. \label{excDY}}
\end{center}
\end{figure}

\vspace*{-4pt}
The nucleon GPDs have been accessed by deeply virtual Compton scattering and deeply virtual meson production with lepton beam. A complementary probe with hadron beam is the exclusive pion-induced DY process. We address the feasibility of measuring the exclusive pion-induced DY process $\pi^- p \to \mu^+\mu^- n$ via a spectrometer at the High Momentum Beamline being constructed at J-PARC in Japan. The experimental signature of the exclusive DY events can be discerned in the missing-mass spectrum of $\mu^+\mu^-$ within 
50 days data taking (Fig.~\ref{excDY}). 

The realization of such measurement at J-PARC will provide a new test of perturbative QCD descriptions of a novel class of hard exclusive reactions as well as offer the possibility of experimentally accessing nucleon GPDs at large time-like virtuality.This measurement will be very interesting to test the leading-twist prediction by Berger, Diehl and Pire~\cite{Berger:2001zn} compared to the prediction that includes transversity GPDs and non-leading twist pion-pole contribution by Goloskokov and 
Kroll~\cite {Goloskokov:2015zsa} which gives about a factor 40 larger cross sections.

%
%
\subsubsection{Spin physics experiments \@ Nuclotron-based Ion Col\-lider fAcility}
\vspace*{-4pt}
\subsubsection*{\hspace*{18pt} \small\em{A.V.~Efremov}}
\vspace*{-4pt}
The PAC of JINR considers the specialized Spin Physics Detector (SPD)~\cite{NICA-SPD} as an essential part of the Nuclotron-based Ion Collider fAcility (NICA) research program and encourages the supporting community to develop this experimental program. This facility would cover a broad range of physics processes such as :
\begin{itemize}
\item nucleon spin structure studies using the DY mechanism; 
\item new nucleon PDFs and J/$\Psi$ production mechanisms; 
\item spin-dependent high-$p_T$ reactions;  
\item direct photon productions in the non-polarized and polarized $pp$ ($pd$) reactions to access gluon distributions in nucleons;
\item spin-dependent effects in elastic $pp$ and $dd$ scattering; 
\item spin-dependent reactions in heavy ion collisions.  
\end{itemize}
Eight asymmetries are to be measured: $A_{LU}$, $A_{UL}$, $A_{TU}$, $A_{UT}$, $A_{LL}$, $A_{TL}$, $A_{LT}$ and $A_{TT}$ which include 23 modulations with amplitudes normalized to unpolarized one. Extraction of different PDFs from these ratios is a task of the global analysis since each single function is a result of convolutions of different PDFs in the quark transverse momentum space. For this purpose one needs either to assume a factorization of the transverse momentum dependence for each PDF, or to transfer them into the impact parameter representation for using Bessel weighted PDFs. The large number of independent functions to be determined from the polarized DY processes at NICA (24 for identical hadrons in the initial state) is sufficient to map out all eight leading twist PDFs for quarks and anti-quarks. Exclusive DY for GPDs measurement could also be considered.
 
The following unpolarized and polarized beams are needed: $pp$, $pd$, $dd$, $pp \uparrow$, $pd \uparrow$, $p \uparrow p \uparrow$, $p \uparrow d \uparrow$, $d \uparrow d \uparrow$. Polarization of both beams will be available at the MultiPurpose Detector (MPD) and the SPD in any possible longitudinal or transverse configurations with absolute values of 50-90\% and long enough beam life time ($\sim$24~h). Measurements of single spin and double spin asymmetries in DY require running in different beam polarization state with  spin flipping every bunch or group of bunches. The beam energy ($\sqrt{s}$) in the $p \uparrow p \uparrow$ mode will range from 12~GeV  up to 27~GeV, and from 4~GeV up to 13.8~GeV in the $d \uparrow d \uparrow$ mode. The expected luminosity are marger than $10^{32}$~cm$^{-2} \cdot s^{-1}$ at 27~GeV in the $pp$ mode, and $10^{30}$~cm$^{-2} \cdot s^{-1}$ at 14~GeV in the $dd$ mode.

In conclusion, feasible schemes of spin manipulations with polarized protons and deuterons at Nuclotron and NICA are suggested. The  final scheme will be settled at a later stage of the project. The development of an international collaboration for the preparation of a proposal and fulfillment of the SPD experiment is strongly encouraged. The Russian government already released a first part of the funding for the construction of the NICA facility, and SPD hall construction is ongoing.

%
%
\subsection{Meson production}

\subsubsection{Hall D}
\vspace*{-4pt}
\subsubsection*{\hspace*{18pt} \small\em{S.~Dobbs}}
\vspace*{-4pt}
GlueX is the flagship experiment in the newly-cons\-truc\-ted Hall D at Jefferson Lab, consisting of an azimuthally-symmetric large-acceptance spectrometer and a linearly polarized photon beam. The first GlueX physics data taking begins this fall, but data from an engineering run in Spring 2016 has been analyzed with the goals of obtaining first physics results and projecting the physics reach for the full approved GlueX running. The first observation of $J/\psi$ production at JLab was shown, with $\sim70$ events in the exclusive reaction $\gamma+p \rightarrow J/\psi + p$ ($J/\psi \rightarrow e^+ e^-$). The data collected in this first physics run will provide roughly 10 times the number of $J/\psi$, which will allow us to study the cross section for $J/\psi$ photo-production for $E_{\gamma} < 12$~GeV, and to probe the possible photo-production of pentaquark resonances. In this initial run, we also expect over 3k Bethe-Heitler events, with $\sim10$ times the number expected for high luminosity GlueX running, which will allow the first studies of TCS at GlueX and open the potential for measuring linearly polarized TCS.  We also expect to make world-class measurements of the time-like form factors of the $\eta$- and $\eta'$-meson.

%

\subsubsection{Near threshold J/$\psi$ photo-production with CLAS12}
\vspace*{-4pt}
\subsubsection*{\hspace*{18pt} \small\em{S.~Stepanyan}}
\vspace*{-4pt}

\begin{figure}[!b]
\begin{center}
\includegraphics[width=0.45\textwidth]{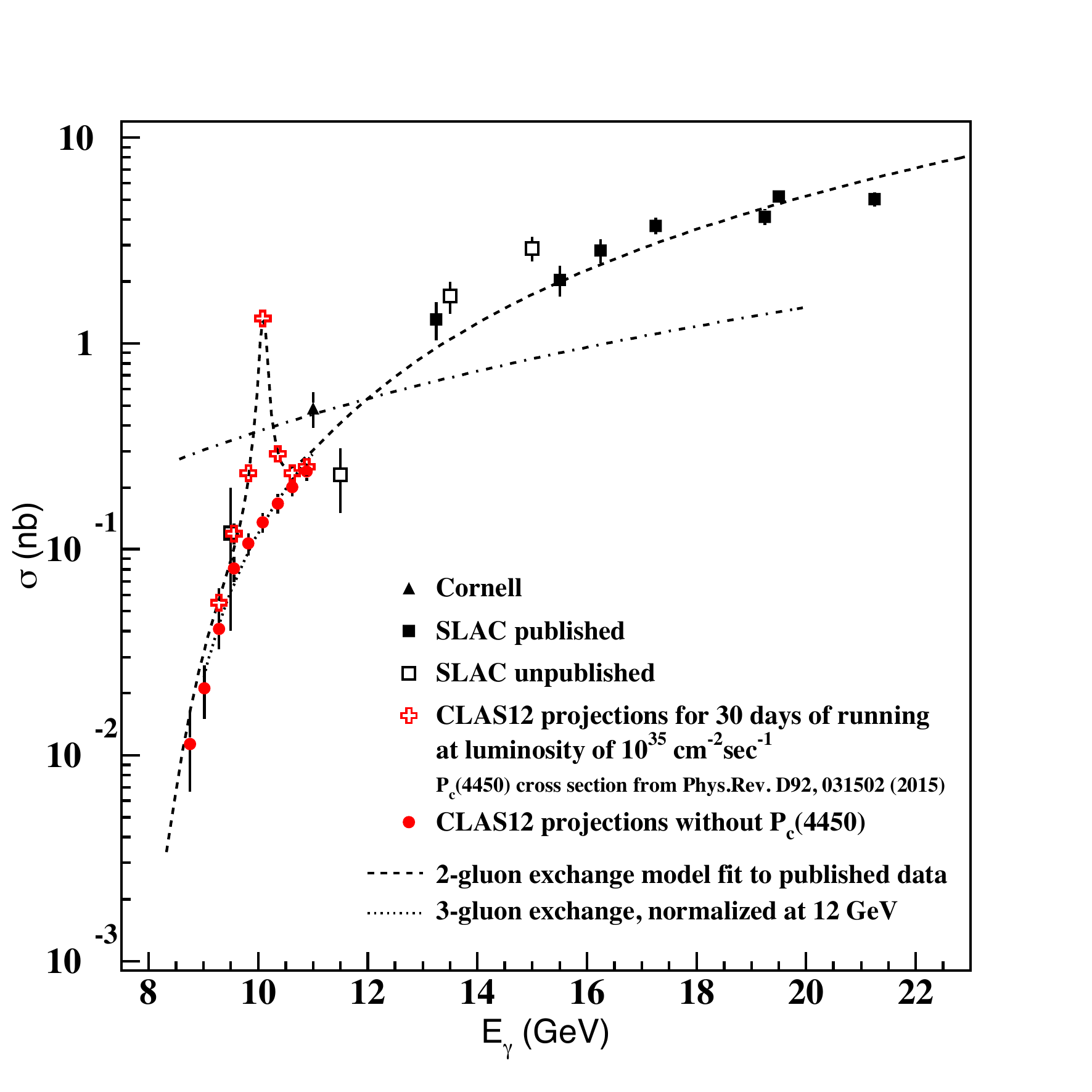}
\caption{Cross section of $J/\psi$ photo-production as a function of photon energy. The cross and bullet points are the expected results from CLAS12 running for 30~days at its design luminosity with an 11~GeV electron beam.}
\label{fig:jp12}
\end{center}
\end{figure}
The large acceptance and high luminosity of CLAS12 together with available high energy beams opens a unique opportunity to explore the gluonic structure of the nucleon by studying charmonium photo-production in the un-measured near threshold region, $E_\gamma < 11$~GeV. The approved experiment with CLAS12~\cite{Nad12} will study near threshold $J/\psi$ photo-production by measuring the differential cross sections as a function of transferred momentum, $t$, from the production threshold ($8.2$~GeV) up to the highest available energy 
($11$~GeV). 

In recent years there has been growing interest in measuring the J/$\psi$-photo-production  cross section close to the threshold. The J/$\psi$ photo-production near threshold provides a direct access to J/$\psi$-$p$ elastic scattering. The elastic scattering with the large transferred momentum to $c\bar c$ pair (near threshold energies) can lead to the dominance of multi-gluon exchange reactions \cite{bchl}, allowing a charmonium bound state formation. As argued in Refs.~\cite{kharz1, Gryniuk:2016mpk} the enhancement of the J/$\psi$ photo-production near threshold could also be due to the dominance of the real part of the scattering amplitude. In \cite{kharz2} the real part of the amplitude is related to the anomalous gluon piece of the energy-momentum tensor, which has been suggested to be one of four components of the proton mass decomposition~\cite{ji}. 

The same measurements access the energy range where the hidden charmed pentaquark states have been found by the LHCb Collaboration~\cite{lhcb}. A variety of models predict sizeable photo-production cross sections for these pentaquarks~\cite{vpk, kr}. CLAS12 will be able not only to perform high precision measurements of $J/\psi$ photo-pro\-duc\-tion near threshold, but also to study photo-production of these pentaquarks. In Fig.~\ref{fig:jp12} the expected results on $J/\psi$ photo-production cross sections as a function of the photon energy are shown without and with the LHCb pentaquark states. In the simulation the lower limit for the $P^+_c(4450)$ cross section from Ref.~\cite{vpk} was used. Consequently, the CLAS12 measurements will shed light on the reaction mechanism near threshold, and provide unique insights into the gluonic structure of the nucleon at large $x$. 

%
\subsubsection{$J/\psi$ photo-production off Nuclei: experimental opportunities at CLAS12}
\vspace*{-4pt}
\subsubsection*{\hspace*{18pt} \small\em{Y.~Ilieva}}
\begin{figure}[ht!]
\begin{center}
\includegraphics[width=0.6\linewidth]{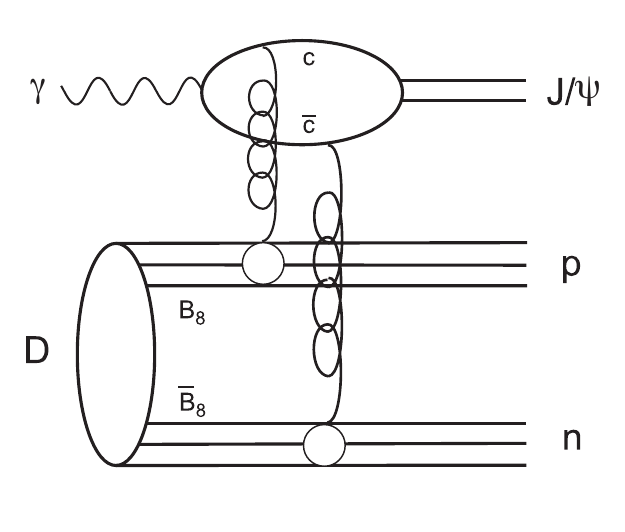}
\caption{\label{fig:hcdiagram}
Schematic diagram of the two-gluon-exchange diagram in incoherent $J/\psi$ photo-production off deuteron. When each exchanged gluon couples to a quark belonging to a different nucleon, the hidden-color component of the deuteron wave function can be probed~\cite{Laget1995}.
}
\end{center}
\end{figure}

\vspace*{-4pt}
Exclusive near-threshold photo-production of $J/\psi$ off the deuteron provides attractive opportunities to study interesting physics. Final-state interactions allow a direct access to the elementary $J/\psi$ cross section~\cite{Freese2013, Lee2013}, while the two-gluon exchange mechanism (Fig.~\ref{fig:hcdiagram}) makes it possible to study aspects of the gluonic structure of the deuteron, such as the deuteron gluonic form factor~\cite{Frankfurt2002}, the deuteron hidden-color component~\cite{Laget1995}, and the gluonic structure of final-state interactions. 

While one measurement of the near-threshold quasi-elastic $J/\psi$ production off the deuteron has been done in the past, data on final-state interactions do not exist. With its high luminosity and large acceptance, the CLAS12 detector offers a window of opportunity to measure near-threshold photo-production of $J/\psi$ off the deuteron~\cite{LOI12-17-001}. The most feasible process is the incoherent production, which can be measured in two ways: (a) untagged real photo-production, where all the final state particles (neutron, proton, and the $e^+e^-$ pair from the $J/\psi$ decay) are detected and the photon is reconstructed via 4-momentum conservation; and (b) quasi-real photo-pro\-duc\-tion, where small $Q^2$ photons are selected by detecting small-angle scattered electrons with the forward tagger, the charged final-state particles are detected in the CLAS12, and the neutron is reconstructed via 4-mo\-men\-tum conservation. Moreover, CLAS12 is perfectly suited to measure coherent photo-production off deuteron and the limitations here arise only from the small reaction cross section. Both, the coherent and the incoherent data will be taken as part of a run-group proposal during an already approved beam time for 90 days with 11 GeV electron beam. The CLAS12 experimental program on heavier nuclei provides opportunities to study $J/\psi$ production off heavier targets, such as C, Fe, and Sn. The $J/\psi$ near-threshold production off nuclei is an extension of the JLab experiment E12-12-001~\cite{Nad12}.

%
%
\subsubsection{Production of charmonium at threshold in Hall A and C at Jefferson Lab}
\vspace*{-4pt}
\subsubsection*{\hspace*{18pt} \em{Z.-E.~Meziani \& S. Joosten}}

\begin{figure}[ht!]
\begin{center}
\includegraphics[angle=0, width=0.45\textwidth]{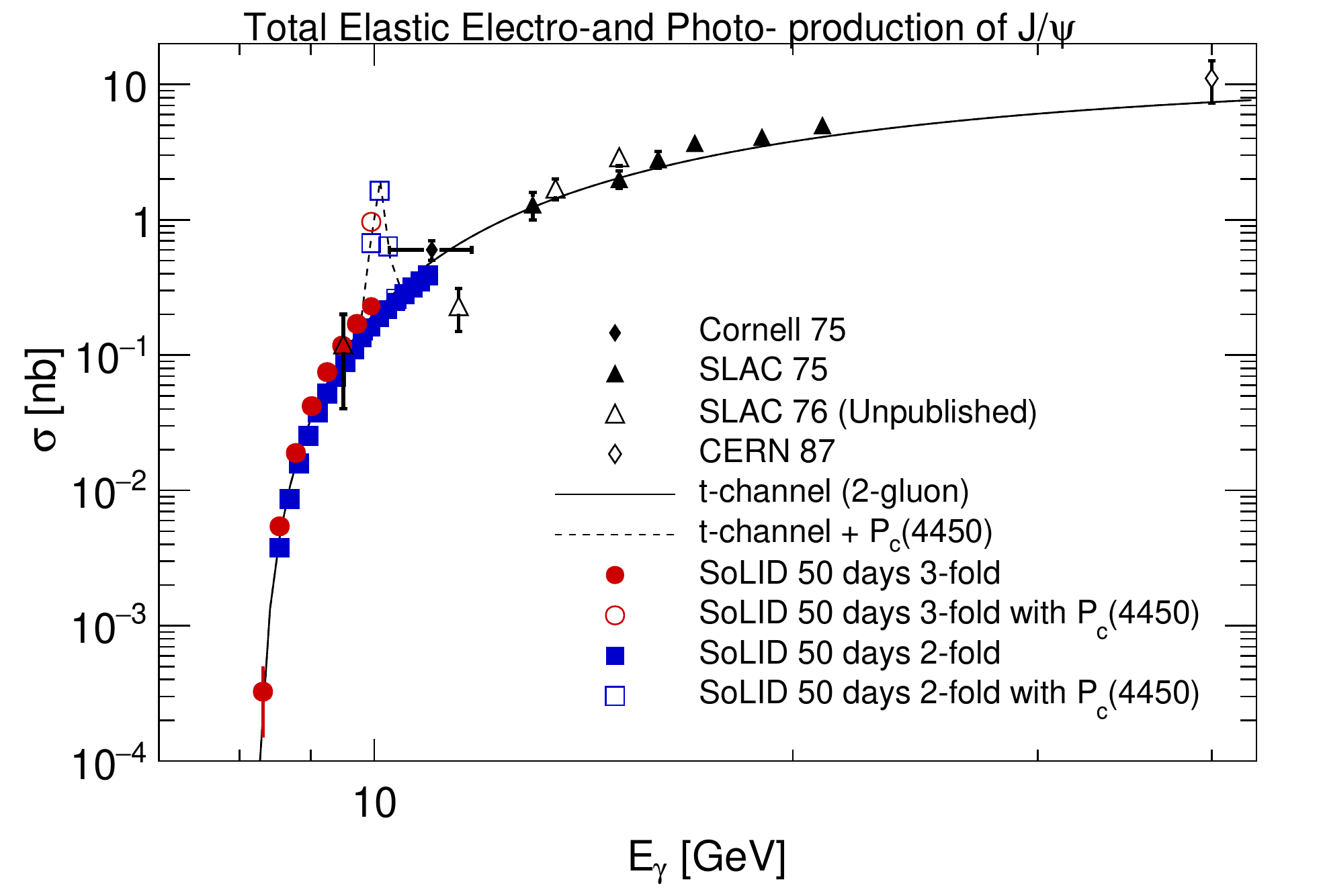}
\caption{Projected uncertainties on the total $J/\psi$ electro- and photo-production cross section. Our projections are based on the 2-gluon exchange model. The central points are positioned at 1.2 times of the predicted total cross section of the 2-gluon  exchange model in order to differentiate our projections from SLAC (unpublished points). SoLID (Fig.~\ref{fig:SoLIDJPsi} ), a high  luminosity device, will offer an unprecedented statistics in photo-production (1627~$J\psi$/day) and electro-production (86~$J\psi$/day) for an approved beam time of 50 days.}
   \label{fig:proj2}
\end{center}
\end{figure}

\begin{figure}[ht!]
\begin{center}
\includegraphics[width=0.45\textwidth]{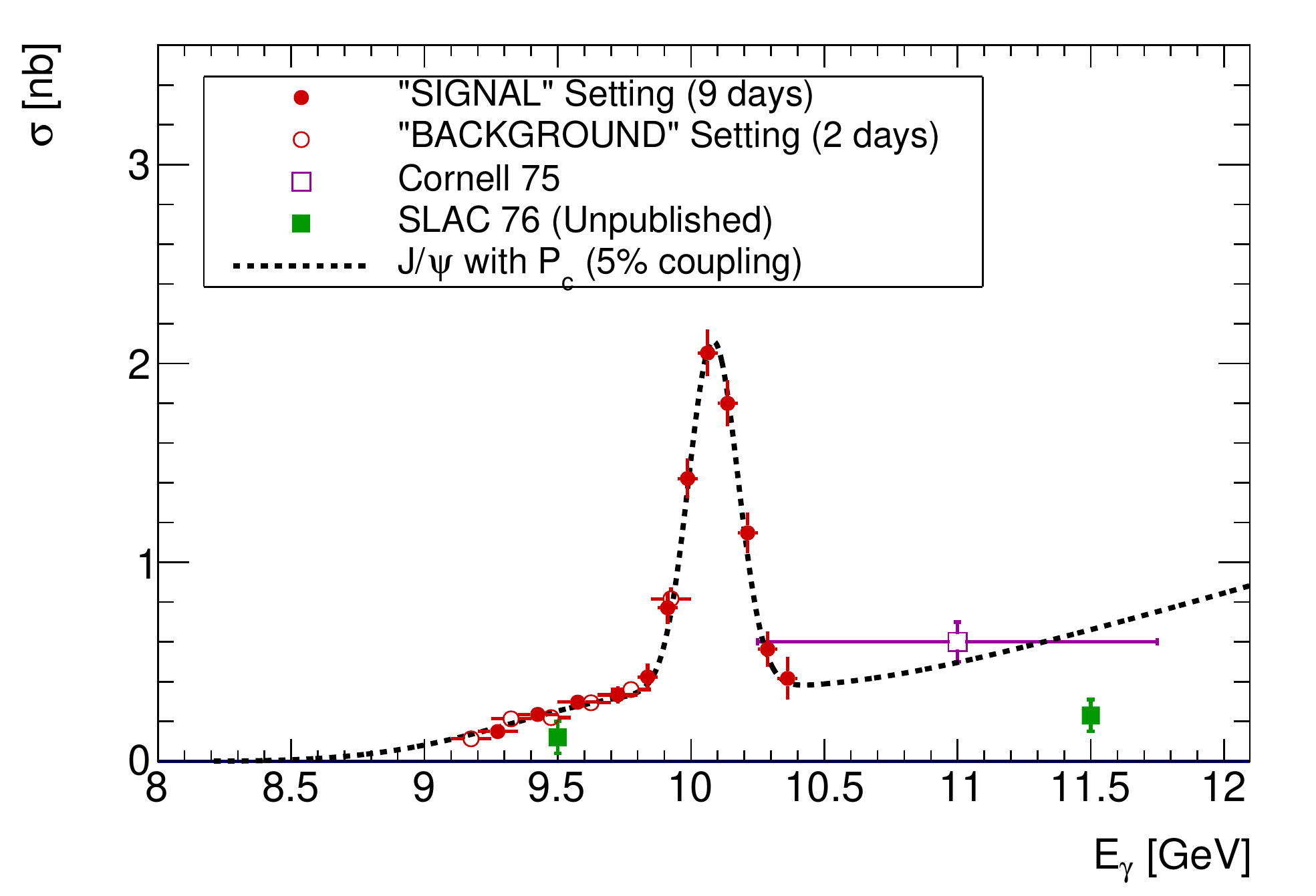}
\caption{Projected impact of this experiment assuming, the (5/2+,~3/2-) case with 5\% coupling, for 9 days of data taking in 
setting \#1 (solid circles) and 2 additional days of data taking in setting \#2 (open circles). The existing data points from 
Cornell and SLAC (unpublished) are also shown.}
\label{fig:impact}
\end{center}
\end{figure}

We present two approved experiments in Hall A~\cite{SoLIDjpsi:proposal} (Fig.~\ref{fig:proj2}) and Hall C~\cite{Meziani:2016lhg})  (Fig.\ref{fig:impact}) at Jefferson Lab that will investigate the pure gluonic component of the QCD strong interaction by measuring the elastic \ensuremath{J/\psi} electro- and photo-production cross section in the threshold region as well  as explore the nature of the recently discovered LHCb charmed pentaquarks. The investigation of the threshold $\ensuremath{J/\psi}$ electro-production at Jefferson Lab is timely with 12 GeV energy upgrade.

The impact of cross section and asymmetry (see Gry\-niuk and Vanderhaeghen in this review article) measurements in the threshold  region will pave the way to answer important questions of QCD predictions in the strong regime. Among them are: What is the strength of the color Van der Waals forces in the nucleon-$\ensuremath{J/\psi}$ system; What is the size of the trace anomaly in the proton mass budget?; What is the nature of the LHCb charmed resonance? The completion of the proposed measurements in tandem with advances in lattice QCD calculations and phenomenology is poised to advance our understanding of QCD, in particular in its much less explored strong interaction regime. 

%
%
\section{Experimental strategy}

%
%
\subsection {Double Deeply Virtual Compton Scattering}

As of today, the DDVCS process is the only reaction enabling investigation of the $x$-dependence separately from the $\xi$-dependence of GPDs (i.e. away from the lines $x=\pm\xi$), particularly the skewness dependence at fixed momentum fraction $x$. It is only at zero-skewness that GPDs acquire a well-defined probability interpretation representing the probability to find a parton carrying the light-cone momentum fraction $x$ of the nucleon at a certain transverse position. DDVCS can therefore constrain the zero-skewness extrapolation required to picture the dynamical distribution of partons inside the nucleon. In absence of relevant polarized positron beams, DDVCS also provides through the first Mellin moment of GPDs an indirect access to the distribution of forces inside the nucleon.

While $e^+e^-$-pair production data will become soon available at CLAS12, it appears that the antisymmetrization of the reaction amplitude significantly affects the experimental signal. Additional contributions from meson decay would further degrade the already small signal such that it is most unlikely that any valuable GPD information could be obtained from $e^+e^-$-pair production with time-like virtual photons. The production of $\mu^+\mu^-$-pair is the only viable and promising channel for DDVCS with an electron beam, or conversely the production of $e^+e^-$-pair with a muon beam possibly at the COMPASS experiment if luminosity allows.

The extraction of CFFs, and consequently GPDs, from experimental data should follow a procedure similar 
to the DVCS case, especially taking into account target mass corrections. These effects on the harmonic decomposition of experimental observables remain to be determined but can be derived straightworardly from the DVCS case by Fourier  transform. In addition, radiative effects on experimental observables should be thoroughly studied and evaluated.

The region of the DDVCS kinematics around $Q^2={Q'}^2$ is of particular interest, because a major and model independent prediction of the handbag formalism is the sign-change of experimental asymmetries as ${Q'}^2$ increases from $Q^2>{Q'}^2$ to $Q^2<{Q'}^2$. This feature is a very strong test of the applicability of the GPD formalism to DDVCS at a given kinematics. Considering the experimental challenges associated to DDVCS in terms of luminosity and detector acceptance, it appears mandatory to establish the pertinence of this process at reachable kinematics at existing facilities. The development of a short scale experiment focused of the sign-change of the DDVCS beam spin asymmetry and measuring its ${Q'}^2$-dependence at fixed $(Q^2, x_B,t)$ kinematics, would assess the appropriate existence of the process and serve the conception and optimization of a large scale DDVCS experimental program in a wide kinematical domain.  

%
%
\subsection {Time-like Compton Scattering}

The TCS and DVCS processes are the exclusive analogs of inclusive DIS with space-like virtual photons and inclusive Drell-Yan with time-like photons~\cite{Die03}. The hard parts of TCS and DVCS are related by analytic continuation which leads in each perturbative order to a precise relationship between them. In that respect, the simultaneous study of TCS and DVCS provides an upmost important test about the  universality of GPDs, which constitutes an essential ingredient of the QCD factorisation approach. 

The experimental demonstration of GPDs universality is the key feature of this program studying the photo-production of $e^+e^-$-pairs. This would consequently enrich the set of experimental observables sensitive to GPDs, contributing to their more accurate determination. Considering intense real or quasi-real photon sources, TCS could even be more competitive than DVCS, as for the determination of the $E$ GPD of the proton because of the limited luminosity capabilities of polarized targets. In addition, similarly to J/$\psi$-prodution, there exists possibilities to study TCS at high energy kinematics in ultra peripheral collisions at the LHC~\cite{LHC-PSW}. However, theoretical predictions require extension to at least the next-to-leading order accuracy since such radiative corrections are known to be potentially large and also to evaluate 
gluons contribution starting only from this order. 

These emphasize the importance of experimental data in the wide kinematical region becoming accessible with the completion of the 12~GeV upgrade at JLab. GPD universality can be experimentally investigated by measuring, for both TCS and DVCS processes, an observable sensitive to a form of the analyticity relationship of the reaction amplitudes. The simplest check related to the Born scattering amplitude is the observation of opposite signs of the imaginary parts of the TCS and DVCS scattering amplitudes, in a straight analogy with a similar sign change of inclusive Sievers distributions for SIDIS  and DY. The already approved experimental program at JLab in halls A and B, and the hall D proposal involving lineraly polarized photons are expected to found the basis of an ambitious experimental program. 

%
%
\subsection {Drell-Yan}

The Drell-Yan process is a powerful tool to explore the partonic structure of nucleons, moving ahead from 1d- to 3d-imaging of their partonic content. Within the framework of the QCD factorisation approach, inclusive DY $N + N^{(\uparrow \downarrow)} \rightarrow l \bar l + X$ is predicted to be sensitive to the transverse polarisation of the target nucleon and to exhibit in accordance single spin asymmetry signals. These signals are generated by a $q \bar{q} g$ twist-3 correlator in the imaginary part of the hadronic tensor. Their magnitude is sensitive to the appearance of the so-called gluonic pole in the parameterisation  expression of this correlator. Similar asymmetry signals  appear in other inclusive DY-type processes as direct photon production $N + N^{(\uparrow \downarrow)} \rightarrow \gamma +q + X$ or the meson (M) induced DY $M +  N^{(\uparrow \downarrow)} \rightarrow \gamma^*(\rightarrow l\bar l) + \bar q + X$, the latter case providing an access to meson distribution amplitudes.

The current experimental programs at COMPASS, Fermilab, PHENIX, J-PARC, and in a near future at NICA, will procure the required experimental data to validate the partonic description of the DY process, and to access GPDs, TDAs, and DAs. New features can potentially be observed, especially in the $\pi^- + p \rightarrow \gamma^*(\rightarrow \mu^+  \mu^-) + n$ experiment at J-PARC where both inclusive and exclusive DY will be simultaneously measured.  

%
%
\subsection {Meson production}

The production of meson is an abundant process which allows to constrain efficiently models and provides access to GPDs and transversity GPDs via the longitudinal and transverse contributions to the cross section, respectively. Furhtermore, the  production of an additionnal photon in the photo-production reaction $\gamma p \rightarrow \gamma M p'$, as a $\gamma \rho$  pair with large invariance mass or a $\gamma \pi$ pair, gives a unique opportunity to access chiral odd GPDs. However,  theoretical improvements are required in the modelling of GPDs in terms of the sensitivity to parton distribution functions, and elaborated profile function to incorporate the $D$-term effects and kinematical corrections for small $Q^2$, small $W$, and large $x$. 

In the kinematical domain where the cross section is not longitudinally dominated, the longitudinal and transverse components of the cross section need to be separated. For instance, the kaon decay of the $\phi$-meson procures a natural  L/T separation. Considering dilepton production, the $J/\Psi$-meson production at threshold is of interest for its unique  sensitivy to the gluon exchange forces.

Meson production experiments at JLab 12 GeV and COMPASS will obtain in a near future an extensive data set to confront with the current understanding of these several reaction channels. This is a mandatory step before investigating the low $x$ physics at the dedicated EIC.

%
%
\section{Conclusion}

Lepton pair production is a very promising tool for deepening our knowledge about nucleon structure. As a time-like partner of DVCS it  enlarges family of processes in which one can test experimentally consequences of the collinear QCD factorisation resulting in particular into universality properties of GPDs. The DDVCS process, involving simultaneously space-like and time-like virtual photons probes, permits the only experimental access to GPD beyond the $x=\pm\xi$ line and allows to test the sign change of the imaginary part of Born scattering  amplitude in the region with a dominant space-like hard scale compared to its time-like counterpart. In addition, the TCS program currently developping at JLab 12 GeV will provide another test of the universality of GPDs. Although it has not been discussed in detail during the workshop, let us add that exclusive dilepton production has also been demonstrated as a tool to access transition distribution amplitudes in nucleon-antinucleon annihilation processes~\cite{TDAPANDA} and feasibility studies have been performed for the PANDA  experiment~\cite{Sing17}, both for the dilepton coming from the $J/\Psi$ and from a virtual photon. Together with the DY and DVMP reactions, the dilepton production channel will allow to assess a more precise and accurate understanding of the partonic structure of hadronic matter.

%
%
\begin{acknowledgement}

We wish to thank Markus Diehl, Dieter Muller, and Bernard Pire for enlighting comments and discussions. This work was supported by the ECT$^{\star}$, the U.S. Department of Energy, Office of Science, Office of Nuclear Physics under contracts DE-AC05-06OR23177 and DE-AC02-76SF00515, and the French Centre National de la Recherche Scientifique. We acknowledge support by the National Science Center in Poland under the contract number 2015/17/B/ST2/01838, by the Polish-French collaboration agreements Polonium and COPIN-IN2P3, by the French ANR PARTONS ANR-12-MONU-0008-01, by the Croatian Science Foundation under the project number 8799, and by the QuantiXLie Center of Excellence KK.01.1.1.01.0004.

\end{acknowledgement}

%
%

%
%

\begin{thebibliography}{999}

\bibitem{Mue94}
  D.~M\"{u}ller, D.~Robaschik, B.~Geyer, F.M.~Dittes, J.~Horejsi, Fortschr. Phys. {\bf 42} (1994) 101.
  
\bibitem{Ji97}
  X.~Ji, Phys. Rev. Lett. {\bf 78} (1997) 610.
  
\bibitem{Rad97}
  A.V.~Radyushkin, Phys. Rev. D {\bf 56} (1997) 5524.

\bibitem{Bur00}
  M.~Burkardt, Phys. Rev. D {\bf 62} (2000) 071503; {\bf 66} (2000) 119903 (Erratum).

\bibitem{Ral02}
  J.P.~Ralston, B.~Pire, Phys. Rev. D {\bf 66} (2002) 111501.
  
\bibitem{Die02}  
  M. Diehl, Eur. Phys. J. C {\bf 25} (2002) 223.

\bibitem{Pol03}
  M.~Polyakov, Phys. Lett. B {\bf 555} (2003) 57.

\bibitem{Hyd11}
  C.E.~Hyde, M.~Guidal, A.V.~Radyushkin, J. Phys. Conf. Ser. {\bf 299} (2011) 012006.

\bibitem{Ber02}
  E.R.~Berger, M.~Diehl, B.~Pire, Eur. Phys. J. C {\bf 23} (2002) 675.
  
\bibitem{Gui03}
  M.~Guidal, M.~Vanderhaeghen, Phys. Rev. Lett. {\bf 90} (2003) 012001. 

\bibitem{Bel03}
  A.V.~Belitsky, D.~M\"{u}ller, Phys. Rev. Lett. {\bf 90} (2003) 022001.
  
\bibitem{Dre70}
  S.~Drell, T.~Yan, Phys. Rev. Lett. {\bf 25} (1970) 316.
  
\bibitem{Gee06}
  (Sea Quest Collaboration) D.F.~Geesaman, P.E.~Reimer et al. FNAL Experiment {\bf E906} (2006). 

\bibitem{Ado16}
  (COMPASS Collaboration) C. Adolph et al. {\bf CERN-EP-2016-250} (2016).

\bibitem{Bel03-1}
  A.V.~Belitsky, D.~M\"{u}ller, Phys. Rev. D {\bf 68} (2003) 116005.

\bibitem{Boe15}
  M.~Boer, M.~Guidal, M.~Vanderhaeghen, Eur. Phys. J. A {\bf 51} (2015) 103.

\bibitem{Fra02}
  L.~Frankfurt, M.~Strikman, Phys. Rev. D {\bf 66} (2002) 031502.
  
\bibitem{Bur03}
  M.~Burkardt, Int. J. Mod. Phys. A {\bf 18} (2003) 173.  

\bibitem{Polyakov:1999gs}
  M.V.~Polyakov, C.~Weiss, Phys.\ Rev.\ D {\bf 60} (1999) 114017.

\bibitem{Teryaev:2005uj}
  O.V.~Teryaev, hep-ph/0510031 (2005).

\bibitem{PSW}
  B.~Pire, L.~Szymanowski, J.~Wagner, Phys. Rev. D {\bf 83} (2011) 034009.
  
\bibitem{MPSzW}  
  D.~M\"{u}ller, B.~Pire, L.~Szymanowski, J.~Wagner, Phys. Rev. D {\bf 86} (2012) 031502.

\bibitem{Braun:2002wu}
  V.M.~Braun, D.Y.~Ivanov, A.~Schafer, L.~Szymanowski, Nucl. Phys. B {\bf 638} (2002) 111.
  
\bibitem{Diehl:2007jb}
  M.~Diehl, D.Y.~Ivanov, Eur. Phys. J. C {\bf 52} (2007) 919.

\bibitem{Brodsky:2008qu} 
  S.J.~Brodsky, F.J.~Llanes-Estrada, A.P.~Szczepaniak, Phys. Rev. D {\bf 79} (2009) 033012.

\bibitem{Brodsky:1972vv} 
  S.J.~Brodsky, F.E.~Close, J.F.~Gunion, Phys. Rev. D {\bf 6} (1972) 177.

\bibitem{Polyakov:2002wz}
  M.V. Polyakov, A.G. Shuvaev, hep-ph/0207153 (2002).

\bibitem{Mueller:2005ed}
  D.~M\"{u}ller, A.~Schafer, Nucl.\ Phys.\ B {\bf 739} (2006) 1.

\bibitem{Muller:2014wxa}
  D.~M\"{u}ller, M.V.~Polyakov, K.M.~Semenov-Tian-Shansky, JHEP {\bf 1503} (2015) 052.

\bibitem{Muller:2015vha}
  D.~M\"{u}ller, K.M.~Semenov-Tian-Shansky, Phys.\ Rev.\ D {\bf 92} (2015) 074025.

\bibitem{GPW}
  A.T.~Goritschnig, B.~Pire, J.~Wagner, Phys. Rev. D {\bf 89} (2014) 094031.

\bibitem{MPSSzW}  
  H.~Moutarde, B.~Pire, F.~Sabati\'e, L.~Szymanowski, J.~Wagner, Phys. Rev. D {\bf 87} (2013) 054029.

\bibitem{IPSzW}
  D.Y.~Ivanov, B.~Pire, L.~Szymanowski, J.~Wagner, EPJ Web Conf. {\bf 112} (2016) 01020.

\bibitem{Anikin:2010wz}
  I.V.~Anikin, O.V.~Teryaev, Phys. Lett. B {\bf 690} (2010) 519; Phys. Lett. B {\bf 751} (2015) 495.

\bibitem{Pivovarov:2015vya}
  A.A.~Pivovarov, O.V.~Teryaev, AIP Conf. Proc. {\bf 1654} (2015) 070008.

\bibitem{Boussarie:2016qop}
  R.~Boussarie, B.~Pire, L.~Szymanowski, S.~Wallon, JHEP {\bf 1702} (2017) 054.

\bibitem{Abb07}
  P.~Abbon {\em et al.} Nucl. Inst. Meth. A {\bf 577} (2007) 455.

\bibitem{Baltz:2007kq} 
  A.J.~Baltz {\it et al.} Phys. Rept. {\bf 458} (2008) 1.

\bibitem{Guzey:2013xba} 
  V.~Guzey, E.~Kryshen, M.~Strikman, M.~Zhalov, Phys. Lett. B {\bf 726} (2013) 290.

\bibitem{Guzey:2013qza} 
  V.~Guzey, M.~Zhalov, JHEP {\bf 1310} (2013) 207 (2013).

\bibitem{Frankfurt:2011cs} 
  L.~Frankfurt, V.~Guzey, M.~Strikman, Phys. Rept. {\bf 512} (2012) 255.

\bibitem{Guzey:2016qwo} 
  V.~Guzey, M.~Strikman, M.~Zhalov, hep-ph/1611.05471.

\bibitem{Alvensleben}
  H.~Alvensleben {\it et al.}, Phys. Rev. Lett. {\bf 30} (1973) 328.

\bibitem{Gryniuk:2015}
  O.~Gryniuk, F.~Hagelstein, V.~Pascalutsa, Phys. Rev. D {\bf 92} (2015) 074031.

\bibitem{Block-Halzen:2004}
  M.M.~Block, F.~Halzen, Phys. Rev. D {\bf 70} (2004) 091901.

\bibitem{Donnachie-Landshoff:2004}
  A.~Donnachie, P.V.~Landshoff, Phys. Lett. B {\bf 595} (2004) 393.

\bibitem{Brodsky:2009gx} 
  S.J.~Brodsky, R.F.~Lebed, Phys. Rev. Lett. {\bf 102} (2009) 213401.
  
\bibitem{Banburski:2012tk} 
  A.~Banburski, P.~Schuster, Phys. Rev. D {\bf 86} (2012) 093007.
  
\bibitem{Brodsky:1997de} 
  S.J.~Brodsky, H.C.~Pauli, S.S.~Pinsky, Phys. Rept. {\bf 301} (1998) 299.
    
\bibitem{Brodsky:2000xy} 
  S.J.~Brodsky, M.~Diehl, D.S.~Hwang, Nucl. Phys. B {\bf 596} (2001) 99.

\bibitem{Penrose:1959vz} 
  R.~Penrose, Proc. Cambridge Phil. Soc. {\bf 55} (1959) 137.

\bibitem{Terrell:1959zz} 
  J.~Terrell, Phys. Rev. {\bf 116} (1959) 1041.

\bibitem{Brodsky:2002cx} 
  S.J.~Brodsky, D.S.~Hwang, I.~Schmidt, Phys. Lett. B {\bf 530} (2002) 99.

\bibitem{Brodsky:2002ue} 
  S.J.~Brodsky, P.~Hoyer, N.~Marchal, S.~Peigne, F.~Sannino, Phys. Rev. D {\bf 65} (2002) 114025.
  
\bibitem{Boer:2002ju} 
  D.~Boer, S.J.~Brodsky, D.S.~Hwang, Phys. Rev. D {\bf 67} (2003) 054003.
  
\bibitem{transGPDacc}
  M.~Diehl {\em et. al.} Phys. Rev. D {\bf 59} (1999) 034023.
  
\bibitem{Coll00}  
  J.C.~Collins {\em et. al.} Phys. Rev. D {\bf 61} (2000) 114015.
  
\bibitem{Ivan02}  
  D.Y.~Ivanov {\em et al.} Phys. Lett. B {\bf 550}(2002) 65.
  
\bibitem{Enbe06}  
  R.~Enberg, B.~Pire, L.~Szymanowski, Eur. Phys. J. C {\bf 47} (2006) 87.
  
\bibitem{Beiy10}  
  M.E.~Beiyad {\em et al.}, Phys. Lett. B {\bf 688} (2010) 154.
  
\bibitem{neutrino}
  B.~Pire, L.~Szymanowski, Phys. Rev. Lett. {\bf 115} (2015) 092001.
  
\bibitem{PSW17}  
  B.~Pire, L.~Szymanowski, J.~Wagner, Phys. Rev. D {\bf 95} (2017) 094001.
  
\bibitem{pion}
  S.~Ahmad, G.R.~Goldstein, S.~Liuti, Phys. Rev. D {\bf 79} (2009) 054014.
  
\bibitem{Golo10}  
  S.V.~Goloskokov, P.~Kroll, Eur. Phys. J. C {\bf 65} (2010) 137; Eur. Phys. J. A {\bf 47} (2011) 112.

\bibitem{Pire05}
  B.~Pire, L.~Szymanowski, Phys. Lett. B {\bf 622}, (2005) 83.
  
\bibitem{Lans12}
  J.-P.~Lansberg, B.~Pire, K.M.~Semenov-Tian-Shansky, L.~Szymanowski, Phys. Rev. D {\bf 86} (2012) 114033; 
  Phys. Rev. D {\bf 87} (2013) 059902 (Erratum).

\bibitem{Pire13}
  B.~Pire, K.M.~Semenov-Tian-Shansky, L.~Szymanowski, Phys. Lett. B {\bf 724} (2013) 99; 
  Phys. Lett. B {\bf 764} (2017) 335 (Erratum).
  
\bibitem{Braun:2011zr} 
  V.M.~Braun, A.N.~Manashov, Phys. Rev. Lett. {\bf 107} (2011) 202001.

\bibitem{Braun:2011dg} 
  V.M.~Braun, A.N.~Manashov, JHEP {\bf 1201} (2012) 085.

\bibitem{Braun:2012bg} 
  V.M.~Braun, A.N.~Manashov, B.~Pirnay, Phys. Rev. D {\bf 86} (2012) 014003.

\bibitem{Braun:2012hq} 
  V.M.~Braun, A.N.~Manashov, B.~Pirnay, Phys. Rev. Lett. {\bf 109} (2012) 242001.

\bibitem{Braun:2014sta} 
  V.M.~Braun, A.N.~Manashov, D.~M\"{u}ller, B.~M.~Pirnay, Phys. Rev. D {\bf 89} (2014) 074022.

\bibitem{Defurne:2015kxq} 
  (Jefferson Lab Hall A Collaboration) M.~Defurne {\it et al.} Phys. Rev. C {\bf 92} (2015) 055202.

\bibitem{Diehl-KrollarXiv:1302.4604}
  M.~Diehl, P.~Kroll, Eur. Phys. J. C {\bf 73} (2013) 2397.

\bibitem{Goloskokov-Krollhep-ph/0611290}
  S.V.~Goloskokov, P.~Kroll, Eur. Phys. J. C {\bf 50} (2007) 829.

\bibitem{Goloskokov-Kroll1106.4897}
  S.V.~Goloskokov, P.~Kroll, Eur. Phys. J. A {\bf 47} (2011) 112.

\bibitem{KrollarXiv:1410.4450}
  P.~Kroll, EPJ Web Conf. {\bf 85} (2015) 01005.

\bibitem{Kumericki:2009uq}
  K.~Kumeri\v{c}ki, D.~M\"{u}ller, Nucl. Phys. B {\bf 841} (2010) 1.

\bibitem{Kumericki:2013br}
  K.~Kumeri\v{c}ki, D.~M\"{u}ller, M.~Murray, Phys. Part. Nucl. {\bf 45} (2004) 723.

\bibitem{Kumericki:2015lhb}
  K.~Kumeri\v{c}ki, D.~M\"{u}ller, EPJ Web Conf. {\bf 112} (2016) 01012.

\bibitem{Gryniuk:2016mpk} 
  O.~Gryniuk, M.~Vanderhaeghen, Phys. Rev. D {\bf 94} (2016) 074001.

\bibitem{Boe15-1}
  M.~Boer, M.~Guidal, J. Phys. G {\bf 42} (2015) 034023.
  
\bibitem{Die03} 
  M.~Diehl, Phys. Rep. {\bf 388} (2013) 41.

\bibitem{Ber15} 
  B.~Berthou {\em et. al.} arXiv:hep-ph/1512.06174 (2015).

\bibitem{Ste16}
  (CLAS Collaboration) S.~Stepanyan {\it et al.} Jefferson Lab Experiment {\bf LOI12-16-004} (2016).

\bibitem{Vou15}
  E.~Voutier {\it et al.} Jefferson Lab Experiment {\bf LOI12-15-005} (2015).

\bibitem{Boe16}
  M.~Boer, M.~Guidal, M.~Vanderhaeghen, Eur. Phys. J. A {\bf 52} (2016) 33.

\bibitem{dual-1} 
  V.~Guzey, T.~Teckentrup, Phys. Rev. D {\bf 79} (2009) 017501. 

\bibitem{dual-2} 
  M.V. Polyakov, K.M. Semenov-Tian-Shansky, Eur. Phys. J. A {\bf 40} (2009) 181198.

\bibitem{double} 
  A.V.~Radyushkin, Phys. Rev. D {\bf 59} (1999) 014030.

\bibitem{Tad15}
  A.~Mkrtchyan {\em et al.} Jefferson Lab Experiment {\bf LOI12-15-007} (2015).
  
\bibitem{Boe15-2}
  M.~Boer, M.~Guidal, M. Vanderhaeghen, arXiv/hep-ph:1501.00270 (2015).

\bibitem{Nad12}
  P.~Nadel-Turonski {\em et al.} Jefferson Lab Experiment {\bf E12-12-001} (2012).

\bibitem{Zha15}
  Z.W.~Zhao {\em et al.} Jefferson Lab Experiment {\bf E12-12-006A} (2015).

\bibitem{Sivers:1989cc}
  D.~W.~Sivers,  Phys.\ Rev.\ D {\bf 41}, 83 (1990).

\bibitem{Collins:2002kn}
  J.~C.~Collins, Phys.\ Lett.\ B {\bf 536}, 43 (2002).

\bibitem{Aghasyan:2017jop}
  (COMPASS Collaboration) M.~Aghasyan {\it et al.} , Phys.\ Rev.\ Lett.\  {\bf 119} (2017) 112002.

\bibitem{Brown:2014sea}
  C.~Brown {\it et al.}, Fermilab Proposal {\bf LOI-1039} (2014).

\bibitem{Aschenauer:2013woa}
  E.~C.~Aschenauer {\it et al.}, arXiv:nucl-ex/1304.0079.

\bibitem{Brodsky:2013oya}
  S.J.~Brodsky, D.S.~Hwang, Y.V.~Kovchegov, I.~Schmidt, M.D.~Sievert, Phys. Rev. D {\bf 88} (2013) 014032.

\bibitem{Berger:2001zn} 
  E.~R.~Berger, M.~Diehl, B.~Pire, Phys. Lett. B {\bf 523} (2001) 265.

\bibitem{Goloskokov:2015zsa} 
  S.~V.~Goloskokov, P.~Kroll, Phys. Lett. B {\bf 748} (2015) 323.

\bibitem{NICA-SPD} 
  I.A.~Savin \textit{et al.} EPJ Web of Conf. {\bf 85} (2015) 02039.

\bibitem{bchl} 
  S.J. Brodsky, E. Chudakov, P. Hoyer, J.M. Laget, Phys. Lett. B {\bf 498} (2001) 23.

\bibitem{kharz1} 
  D. Kharzeev et al. Eur. Phys. J. {\bf C 9} (1999) 459.

\bibitem{kharz2} 
  D. Kharzaeev, arXiv:nucl-th/9601029 (1996).

\bibitem{ji} 
  X. Ji, Phys. Rev. Lett. {\bf 74} (1995) 1071.

\bibitem{lhcb} 
  R. Aaij {\it et al.} {\it (LHCb Collaboration)}, Phys. Rev. Lett. {\bf 115} (2015) 072001.

\bibitem{vpk} 
  V. Kubarovsky, M.B. Voloshin, Phys.Rev. D {\bf 92} (2015) 031502.

\bibitem{kr} 
  M. Karliner, J.L. Rosner, Phys. Lett. B {\bf 752} (2016) 329.

\bibitem{Freese2013} 
  A.J.~Freese, M.M.~Sargsian, Phys. Rev. C {\bf 88} (2013) 044604.
  
\bibitem{Lee2013} 
  J.J.~Wu, T.-S.~H.~Lee, Phys. Rev. C {\bf 88} (2013) 015205.
  
\bibitem{Frankfurt2002} 
  L.~Frankfurt, M.~Strikman, Phys. Rev. D {\bf 66} (2002) 031502(R).
  
\bibitem{Laget1995} 
  J.M.~Laget, R.~Mendez-Galain, Nucl. Phys. A {\bf 581} (1995) 397.
  
\bibitem{LOI12-17-001} 
  Y.~Ilieva \textit{et al.} Jefferson Lab Letter-of-Intent {\bf LOI12-17-001} (2017).
  
\bibitem{SoLIDjpsi:proposal}
  Z.-E.~Meziani {\it et al.} Jefferson Lab Experiment {\bf E12-12-006} (2012).
  
\bibitem{Meziani:2016lhg}
  Z.~E. Meziani {\em et~al.} Jefferson Lab Experiment {\bf E12-16-007} (2016).

\bibitem{LHC-PSW}
  B.~Pire, L.~Szymanowski, J.~Wagner, Phys. Rev. D {\bf 79} (2009) 014010.

\bibitem{TDAPANDA}
  (PANDA Collaboration) B.P.~Singh {\it et al.} Eur. Phys. J. A {\bf 51} (2015) 107.
  
\bibitem{Sing17}  
  (PANDA Collaboration) B.P.~Singh {\it et al.} Phys. Rev. D {\bf 95} (2017) 032003.
    
\end{thebibliography}
\end{document}